\documentclass[11pt,twoside]{article}

\usepackage[usenames,x11names,table]{xcolor}
\usepackage{float}
\usepackage[font={rm,md,sl},format=hang]{subfig}
\usepackage[font={rm,small,sl}]{caption}
\usepackage{graphicx}
\usepackage[pagebackref,breaklinks]{hyperref}
\usepackage{breakurl}
\usepackage[margin=1.1in]{geometry}
\usepackage{enumerate}
\usepackage{tikz}
\newcommand*\circled[1]{\tikz[baseline=(char.base)]{
            \node[shape=circle,draw,inner sep=0.6pt] (char) {#1};}}
\usepackage{listings}

\hypersetup{%
	pdfusetitle=true,%
	bookmarks=true,%
	pdftitle={Accelerating Analytical Processing in MVCC using Fine-Granular High-Frequency Virtual Snapshotting},%
    pdfauthor={Ankur Sharma, Felix Martin Schuhknecht, Jens Dittrich},     
    pdfsubject={Transactional Processing, Snapshotting},   
    pdfcreator={Ankur Sharma, Felix Martin Schuhknecht, Jens Dittrich},   
    pdfproducer={\LaTeX}, 
	pdfkeywords={ransactional Processing, Snapshotting, OLTP, OLAP, Virtual Memory},%
	pdfdisplaydoctitle=true,%
	pdfstartview=Fit,%
	colorlinks=false%
}

\renewcommand*{\backrefalt}[4]{%
\ifcase #1 %
No citations.%
\or
One citation on page #2.%
\else
#3 citations on pages #2.%
\fi
}

\colorlet{color}{cyan!50}
\newcommand{\smalltt}[1]{{\texttt{\small #1}}}

\newcolumntype{L}[1]{>{\raggedright\let\newline\\\arraybackslash\hspace{0pt}}m{#1}}
\newcolumntype{C}[1]{>{\centering\let\newline\\\arraybackslash\hspace{0pt}}m{#1}}
\newcolumntype{R}[1]{>{\raggedleft\let\newline\\\arraybackslash\hspace{0pt}}m{#1}}

\begin{document}


\title{Accelerating Analytical Processing in MVCC using Fine-Granular High-Frequency Virtual Snapshotting}


\author{Ankur Sharma\qquad Felix Martin Schuhknecht\qquad Jens Dittrich\\
\\
Information Systems Group\\
Saarland Informatics Campus\\
\url{http://infosys.uni-saarland.de}}

\date{\today}

\maketitle

\begin{abstract}

Efficient transactional management is a delicate task. As systems face transactions of inherently different types, ranging from point updates to long running analytical computations, it is hard to satisfy their individual requirements with a single processing component. 
Unfortunately, most systems nowadays rely on such a single component that implements its parallelism using multi-version concurrency control (MVCC). While MVCC parallelizes short-running OLTP transactions very well, it struggles in the presence of mixed workloads containing long-running scan-centric OLAP queries, as scans have to work their way through large amounts of versioned data.
To overcome this problem, we propose a system, which reintroduces the concept of heterogeneous transaction processing: OLAP transactions are outsourced to run on separate (virtual) snapshots while OLTP transactions run on the most recent representation of the database. Inside both components, MVCC ensures a high degree of concurrency.

The biggest challenge of such a heterogeneous approach is to generate the snapshots at a high frequency. Previous approaches heavily suffered from the tremendous cost of snapshot creation. In our system, we overcome the restrictions of the OS by introducing a custom system call \smalltt{vm\_snapshot}, that is hand-tailored to our precise needs: it allows fine-granular snapshot creation at very high frequencies, rendering the snapshot creation phase orders of magnitudes faster than state-of-the-art approaches.
Our experimental evaluation on a heterogeneous workload based on TPC-H transactions and handcrafted OLTP transactions shows that our system enables significantly higher analytical transaction throughputs on mixed workloads than homogeneous approaches.

In this sense, we introduce a system that accelerates \textit{Analytical} processing by introducing custom \textit{Kernel} functionalities: \textit{AnKerDB}.

\end{abstract}

\section{Introduction}
\label{sec:introduction}

Fast concurrent transactional processing is one of the major design goals of basically every modern database management system. To fully utilize the large amount of hardware parallelization that is nowadays available even in commodity servers, the right concurrency control mechanism must be chosen. 

Interestingly, a large number of database systems, including major players like PostgreSQL~\cite{postgres}, Microsoft Hekaton~\cite{hekaton}, SAP HANA~\cite{hana}, and HyPer~\cite{hypermvcc}, currently implement a form of multi-version concurrency control (MVCC)~\cite{pavlo_mvcc, bernstein_concurrency} to manage their transactions. It allows a high degree of parallelism as reads do not block writes. The core principle is straight-forward: if a value is updated, its old version is not simply replaced by the new version. Instead, the new version is stored alongside with the old one in a version chain, such that the old version is still available for reads that require it. Timestamps ensure that transactions access only the version they are allowed to see.

\subsection{Limitations of Classical MVCC}
\label{ssec:limitations}
In classical MVCC implementations, all transactions, no matter whether they are short running OLTP transactions or scan-heavy OLAP transactions, are treated equally and are executed on the same (versioned) database. 
While this form of \textit{homogeneous processing} unifies the way of transaction management, it also has a few unpleasant downsides under mixed workloads:

First and foremost, scan-heavy OLAP transactions heavily suffer when they have to deal with a large number of lengthy version chains. During a scan, each of these version chains must be traversed to locate the most recent version of each item that is visible to the transaction. This involves expensive timestamp comparisons as well as random accesses when going through the version chains that are typically organized as linked lists. As scans typically take time, a large amount of OLTP transactions can perform updates in parallel and lengthy version chains built up during the execution.

Apart from this, these version chains must be garbage collected from time to time to remove versions that can not be seen by any transaction in the system. Under classical MVCC, this is typically done by a separate cleanup thread, which frequently traverses all present chains to locate and to delete outdated versions. This thread has to be managed and synchronized with the transaction processing, utilizing precious resources. 

Obviously, the mentioned problems are directly connected to the processing of scan-heavy OLAP transactions in the presence of short-running modifying OLTP transactions. Such a \textit{heterogeneous workload}, consisting of transactions of inherently different nature, simply does not fit to \textit{homogeneous processing}, which treats all incoming transactions in the same way. Unfortunately, such a homogeneous processing model is used in the state-of-the-art MVCC systems.

\subsection{Heterogeneous Processing}
\label{ssec:het_processing}

But why exactly do state-of-the-art systems rely on a homogeneous processing model, although it does not fit to the faced workload? Why don't they implement \textit{heterogeneous processing}, which classifies transactions based on the type and executes them in separation?

To answer these questions, let us look at the development of the prominent HyPer\cite{hyper, hypermvcc} system. Early versions of HyPer actually implemented heterogeneous processing~\cite{hyper}: the transactions were classified into the categories OLTP and OLAP and consequently executed on separate representations of the database. The short running modifying OLTP transactions were executed on the most recent version of the data while long-running OLAP transactions were outsourced to run on \textit{snapshots}. These snapshots were created from time to time on the up-to-date version of the database. 

While this concept mapped the mixed workload to the processing system in a very natural way, the engineers faced a severe problem: the creation of snapshots turned out to be very expensive~\cite{hypermvcc}. 
To snapshot, HyPer utilized the \smalltt{fork}~system call. This system call creates a child process that shares its virtual memory with the one of the parent process. Both processes perform copy-on-write to keep changes locally, thus implementing a form of (virtual) snapshotting.
While this principle is cheaper than physical snapshotting, forking processes is still costly. Thus, the engineers were forced to move away from heterogeneous processing to a homogeneous model, fully relying on MVCC in their current version.

\subsection{Challenges}

Despite the challenges one has to face when implementing a heterogeneous model, we believe it is the right choice after all. Matching the processing system to the workload is crucial for performance. This is exactly the goal of our main-memory transaction processing system coined~\textit{AnKerDB}, which we will propose in the following. Still, to do so, we have to discuss two problems first:
\begin{enumerate}[(a)]
\item Obviously, MVCC is the state-of-the-art concurrency control mechanism in main-memory systems. In AnkerDB, we intend to apply it as well. But how to combine state-of-the-art MVCC with a heterogeneous processing model? 
\item Apparently, state-of-the-art snapshotting mechanisms are not capable of powering a heterogeneous processing model. How to realize a fast snapshotting mechanism, that allows the creation of snapshots at a high frequency and at fine granularity? 
\end{enumerate}

\noindent Let us discuss these questions one by one in the following.

\subsubsection{MVCC in Heterogeneous Processing}
Classical systems implement MVCC in a homogeneous processing model, where all transactions are treated equally and executed on the same versioned database.
In contrast to that, in AnKerDB we want to extend the capabilities of MVCC by reintroducing the concept of \textit{heterogeneous processing}, where incoming transactions are classified by their type and treated independently. By this, we are able to utilize the advantages of MVCC \textit{while} avoiding its downsides. 

The concept works as follows: based on the classification, we separate the short-running OLTP transactions from the long-running (read-only) OLAP transactions. 
Conceptually, the modifying OLTP transactions run concurrently on the most recent version of the database and build up version chains as in classical MVCC. In parallel, we outsource the read-only OLAP transactions to run on separate (read-only) snapshots of the versioned database.

These snapshots are created at a very high frequency to ensure freshness. Thus, instead of dealing with a single representation of the database that suffers from a large number of lengthy version chains, as it is the case in systems that rely on homogeneous processing, we maintain a most recent representation inside of an OLTP component alongside with a set of snapshots, which are present in the OLAP component.
Naturally, each of the representations contains fewer and shorter version chains, which largely reduces the main problem described in Section~\ref{ssec:limitations}.

Apart from that, using snapshots has the pleasant side-effect that the garbage collection of version chains becomes extremely simple: We remove the chains automatically with the deletion of the corresponding snapshot, if it can not be seen by any transaction anymore. Other systems like PostgreSQL have to rely on a fine-granular garbage collection mechanism for shortening the version-chains, requiring precious resources. 
In contrast to that, by using snapshotting, we are able to solve the problem of complex garbage collection techniques implicitly. 

\subsubsection{High-Frequency Snapshotting}
With the high-level design of the heterogeneous processing model at hand, it remains the question how to realize efficient snapshotting.
The approach stands and falls with the ability to generate snapshots at a \textit{very high frequency} to ensure that transactions running on the snapshots have to deal only with few and short version chains. In this regard, previous approaches that relied on snapshotting suffered under the expensive snapshot creation phase and consequently moved away from snapshotting. As mentioned, early versions of HyPer~\cite{hyper}, which also used a heterogeneous processing model, created virtual snapshots using the system call~\smalltt{fork}. 
This call is used to spawn child processes which share their entire virtual memory with the parent process. The copy-on-write mechanism, that is carried out by the operating system on the level of memory pages ensures that changes remain local in the related processes. While this mechanism obviously implements a form of snapshotting, process forking is very expensive. Thus, it is not an option for our case as we require a more lightweight snapshotting mechanism.

In our recent publication on the \textit{rewiring}~\cite{rewiring} of virtual memory, we already looked into the case of snapshot creation. With rewiring, we are able to manipulate the mapping from virtual to physical memory pages at runtime in user space. In~\cite{rewiring}, we used this technique to snapshot an existing virtual memory area~$v_1$, which maps to a physical memory area~$p_1$, by manually establishing a mapping of a new virtual memory area~$v_2$ to $p_1$. While this approach is already significantly faster than using~\smalltt{fork} as we stay inside a single process, it is still not optimal as the mapping must be reconstructed page-wise in the worst case --- a costly process for large mappings as individual system calls must be carried out.

Unfortunately, all the existing solutions are not sufficient for our requirements on snapshot creation speed.  
Therefore, in AnKerDB, we implement a more sophisticated form of \textit{virtual} snapshotting. We do not limit ourselves by using the given general purpose system calls. Instead, we introduce our own custom system call coined~\smalltt{vm\_snapshot} and integrate the concept of rewiring~\cite{rewiring} directly into the kernel. Using our call, we can essentially snapshot arbitrary virtual memory areas within a single process at any point in time. The virtual snapshots share their physical memory until a write to a virtual page happens, which creates a local physical page. This allows us to create snapshots with a small memory footprint at a very low cost, allowing us to build them at a high frequency. Consequently, the individual snapshots contain few and short version chains and enable efficient scans.  

\subsection{Structure \& Contributions}

Before we start with the detailed presentation of the system design and the individual components, let us outline the contributions we make in the following work:\\

\noindent \textbf{(I)} We present AnKerDB, a prototypical main-memory (column-oriented) transaction processing system, which supports the efficient concurrent execution of transactions under a \textit{heterogeneous processing model} under \textit{full serializability guarantees}. Short-running (modifying) transactions concurrently run on the most recent version of the data using MVCC. Meanwhile long-running read-only transactions run on (versioned) snapshots in parallel.

\noindent \textbf{(II)} We realize the snapshots in form of \textit{virtual snapshots} and heavily accelerate the snapshotting process by introducing a \textit{custom system call} coined \smalltt{vm\_snapshot} to the Linux kernel. This call directly manipulates the virtual memory subsystem of the OS and allows a significantly higher snapshotting frequency than state-of-the-art techniques. We demonstrate the capabilities of \smalltt{vm\_snapshot} in a set of micro-benchmarks and compare it against the existing physical and virtual snapshotting methods.

\noindent \textbf{(III)} We create snapshots on the \textit{granularity of a column}, instead of snapshotting the entire table or database as a whole. This is possible due to the flexibility of our custom system call~\smalltt{vm\_snapshot}. Therefore, we are able to limit the snapshotting effort to those columns, which are actually accessed by the incoming transactions.

\noindent \textbf{(IV)} We create \textit{snapshots of versioned columns} to keep the snapshot creation phase as cheap as possible. To create a snapshot, the current column is virtually snapshotted using our custom system call \smalltt{vm\_snapshot} and the current version chains are handed over. Running transactions can still access all required versions from the fresh snapshot. As the snapshot is read-only, all further updates happen to the up-to-date column, creating new version chains. As a side-effect, we avoid any expensive garbage collection mechanism as dropping an old snapshot drops all old version chains with it.

\noindent \textbf{(V)} We perform an extensive experimental evaluation of AnKerDB.  First, we compare our heterogeneous transactions processing model with classical homogeneous MVCC under both snapshot isolation and full serializability guarantees, executing mixed OLTP/OLAP workloads based on TPC-H queries and hand-tailored OLTP transactions. To enable this form of evaluation, AnKerDB can be configured to support both heterogeneous and homogeneous processing (by disabling snapshotting) as well as the required isolation levels. We show that our approach offers a drastically higher transaction throughput under mixed workloads. 

\noindent The paper is structured in the following way: In Section~\ref{sec:ankerdb}, we describe the heterogeneous design of AnKerDB and motivate it with the problems of state-of-the-art MVCC approaches. As the heterogeneous design requires a fast snapshotting mechanism, we discuss the currently available snapshotting techniques to understand their strengths and weaknesses in Section~\ref{sec:snapshotting}. As a consequence, in Section~\ref{sec:rewiringplusplus}, we propose our own snapshotting method based on our custom system call~\smalltt{vm\_snapshot}.
Finally, in Section~\ref{sec:experimental_evaluation}, we evaluate AnKerDB in different configurations and show the superiority of heterogeneous processing.

\section{AnKer DB}
\label{sec:ankerdb}

As already outlined, the central component of AnKerDB is a heterogeneous processing model, which separates OLTP from OLAP processing using virtual snapshotting. Both in the up-to-date representation of the data as well as in the snapshots, we want to use MVCC as the concurrency control mechanism. To understand our hybrid design, let us first see how MVCC is working within a single component.

\subsection{Classical MVCC}
\label{ssec:mvcc}

To understand the mechanisms of classical MVCC, let us go through the individual components. Initially, the data is unversioned and present in the column. Thus, there exist no version chains. If a transaction updates an entry, we first store the new value locally inside the local memory space of the transaction. When the commit is carried out, the update has to be materialized in the column. To do so, the old value is stored in the (freshly created) version chain of that row and the old value is overwritten with the new one in the column in-place. Thus, we store the versions in a \textit{newest-to-oldest} order. 
Other systems as e.g. HyPer~\cite{hypermvcc} rely on this order as well as it favors younger transactions: they will find their version early on during the chain traversal. 
Obviously, the version chains can become arbitrarily long, if consecutive updates to the same entry happen. Along with the version, we store a unique timestamp of the update that created that version. This is necessary to ensure that transactions, that started before the (committed) update happened, do not see the new version of the entry but still the old one. Unfortunately, reading a versioned column can also become arbitrarily expensive: for every entry that a transactions intends to read and that has a version chain, the chain must be traversed under comparisons of the timestamps to locate the proper version.
In summary, if a large amount of lengthy version chains is present and a transaction intends to read many entries, version chain traversal cost becomes significant.

Besides the way of versioning the data, the guaranteed isolation level is an important aspect in MVCC. As a consequence of its design, MVCC implements snapshot isolation guarantees by default. During its lifetime, a transaction~$T$ sees the committed state of the database, that was present at $T$'s start time. The updates of newer transactions, which committed during $T$'s lifetime, are not seen by $T$. Write-write conflicts are detected at commit time: if $T$ wants to write to an entry, to which a newer committed transactions already wrote, $T$ aborts. 
Still, under snapshot isolation, so called write-skew anomalies are possible. 

Fortunately, MVCC can be extended to support full serializability. To do so, we extend the commit phase of a transaction with additional checks. If a transactions~$T$ wants to commit, it validates its read-set by inspecting if any other transaction, that committed during $T$'s lifetime changed an entry in a way that would have influenced $T$'s result. If this is the case, $T$ has to abort as its execution was based on stale reads. To perform the validation, we adopt the efficient approach applied in HyPer~\cite{hypermvcc}, which is again based on the technique of precision locking~\cite{weikum_precision_locking}. Essentially, we track the predicate ranges on which the transaction filtered the query result. During validation, it is checked whether any write of any recently committed transaction intersects with the predicate ranges. If an intersection is identified, the transaction aborts.

\subsection{Heterogeneous MVCC}
\begin{figure}[!htb]
	\vspace{-0.2cm}
	\begin{center}
		\includegraphics[width=6.5cm, height=19cm]{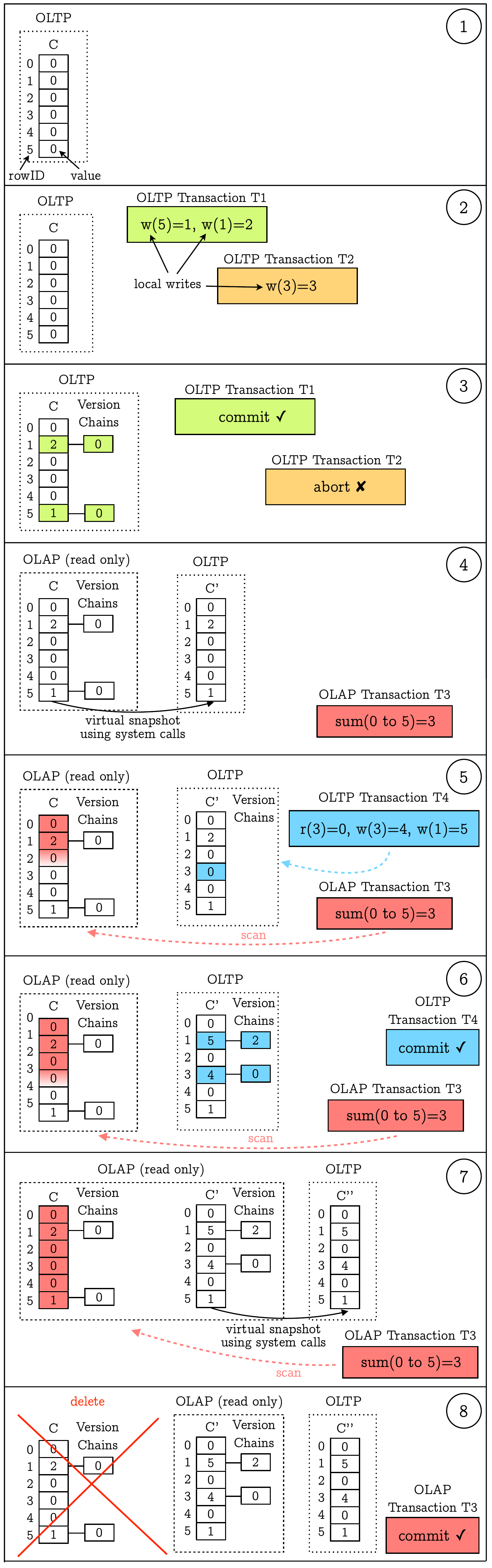}
	\end{center}
	\vspace{-0.5cm}
	\caption{Example of the \textbf{heterogeneous processing model} in AnKerDB.}
	\label{figs:ankerdb_concept}
	\vspace{-0.3cm}
\end{figure}

To overcome the aforementioned limitations of classical MVCC implementations, we realize a heterogeneous transaction processing model in AnKerDB.
Two components are present side by side: one component is responsible for the concurrent processing of short-running transactions (coined \textit{OLTP component} in the following), while the other one can perform long-running read-only transactions in parallel (coined \textit{OLAP component} from here on). Incoming transactions are classified into being either an OLTP or an OLAP transaction and send to the respective component for processing. The challenge is to combine the concept of heterogeneous processing with MVCC. Let us look at the components in detail at the case of an example we show in Figure~\ref{figs:ankerdb_concept}. 

\subsubsection{Example}

Step~\circled{1}: For the following discussion, we assume that our table consists of a single column~$C$ of $6$~rows (identified by row~$0$ to $5$), which all contain the value $0$ in the beginning. This column~$C$ is located in the OLTP component and represents the up-to-date representation of the column. Since there are no snapshots present yet, the OLAP component essentially does not exist. 
 
Step~\circled{2}: Two OLTP transactions $T_1$ and $T_2$ arrive and intend to perform a set of writes. The first write of $T_1$ ($w(5)=1$) intends to update at row~$5$ the value $0$ with the new value~$1$. However, instead of replacing the old value in the column with the new value in-place, we store the new value locally inside the transaction~$T_1$ and keep the column untouched as long as the transaction does not commit. 
In the same fashion, the remaining write of $T_1$ ($w(1)=2$) as well the write of $T_2$ ($w(3)=3$) are performed only locally inside their respective transactions. Note that all three written values are uncommitted so far and can only be seen by the transactions that performed the respective writes.

Step~\circled{3}: Let us now assume that $T_1$ commits while $T_2$ intentionally aborts. The commit of $T_1$ now actually replaces in column~$C$ at row~$5$ the old value~$0$ with the new value~$1$. Of course, the old value~$0$ is not discarded, but stored in a newly created version chain for that row. The same procedure is performed at row~$1$ where the old value~$0$ is replaced with the new value~$2$, moving the old value into the version chain. Note that we implement a timestamp mechanism (logging both the start and end time of a transactions commit phase) to ensure that both writes of $T_1$ becomes visible atomically to other transactions. 
As no other transactions modified row $1$ and $5$ during the lifetime of $T_1$, the commit succeeds and satisfies full serializability, that we guarantee for all transactions. In contrast to that, the abort of $T_2$ simply discards the local change of row~$3$. This strategy makes aborts very cheap, as no rollback must be performed.

Step~\circled{4}: An OLAP transaction~$T_3$ arrives, which intends to scan and sum up the values of all rows of the column, denoted by sum($0$~to~$5$). As no snapshot is present yet to run $T_3$ on, the first snapshot is taken. Using our custom system call (which will be described in Section~\ref{sec:rewiringplusplus} in detail), we snapshot the column~$C$, resulting in a (virtual) duplicate of the column in form of~$C'$. It is important to understand that this duplicate~$C'$ will become the most recent version of the column in the OLTP component. The ``old" column~$C$ along with its build-up version chains is logically moved to the OLAP component and becomes read-only.

Step~\circled{5}: Another OLTP transactions~$T_4$ arrives, that intends to perform a read~$r(3)$ followed by two writes ($w(3)=4$, $w(1)=5$). The read~$r(3)$ is simply performed by accessing the current value of row~$3$ of the representation in the OLTP view, resulting in $r(3)=0$. The two successive writes are stored locally inside~$T_4$ and are not visible for other transactions.  
In parallel to the depicted operations of~$T_4$, our OLAP transaction~$T_3$, which sums up the column values, starts executing in the OLAP component on $C$. 
As the snapshot is older than $T_3$, it can simply scan the column~$C$ without inspecting the version chains.  

Step~\circled{6}: While the scan of $T_3$ is running, $T_4$ decides to commit. This commit does not conflict with the execution of $T_3$ in any way, as the transactions run in different components. The local writes $w(3)=4$ and $w(1)=5$ are materialized in~$C'$ and the old versions are again stored in version chains. 

Step~\circled{7}: Another snapshot is taken to have a more up-to-date representation ready for incoming OLAP transactions. Again we use our custom system call and snapshot the column~$C'$ that is located in the OLTP component, resulting in a (virtual) duplicate of the column in form of~$C''$. As before, the roles are changed: The new duplicate~$C''$ becomes the most recent representation of the column in the OLTP component, while~$C'$ with its version chains is moved to the OLAP component. Note that both $C'$ as well as $C$ are now present in the OLAP component side by side, with $T_3$ still running on~$C$. 

Step~\circled{8}: The OLAP transaction~$T_3$ finishes its scan and commits. This makes~$C$ obsolete, as a newer representation~$C'$ already exists. As no transaction is running, we can safely delete the oldest snapshot~$C$, as no incoming transaction can access any of its versions anymore.

\subsubsection{Snapshot Synchronization}
\label{ssec:sync}

For simplicity, in the previous example all transactions worked solely on a single column. However, a database usually consists of several tables, each equipped with a large number of attributes and therefore, some form of \textit{snapshot synchronization} is necessary. In this context, snapshot synchronization means that a transaction, which accesses multiple columns, has to see all columns consistent with respect to a single point in time. A trivial way of achieving this is to simply snapshot all columns of all tables when a snapshot is requested. However, this causes unnecessary overhead as we might access only a small subset of the attributes. 
Therefore, in AnKerDB, we implement a lazy approach: when a snapshot creation is triggered, only a timestamp for that snapshot is logged and no actual snapshotting is performed yet. If a transaction comes in, that accesses a set of columns, it is checked whether there are snapshots present for these columns. If not, they are materialized. This ensures that columns, which are never touched are never materialized as snapshots. 

\subsubsection{Snapshot Consistency}
\label{ssec:consistency}

In the previous example, we simply created a snapshot when the individual OLAP transactions required it. In our actual implementation, we trigger a snapshot creation after $n$~commits happened to the database. When this happens and the previously described access triggers the actual materialization of the snapshot using our system call, we have to ensure that no other transactions modify the column while the snapshot is under creation. We ensure this using a shared lock on the column, which must be acquired by any transaction to update. When materializing a snapshot, an exclusive lock must be acquired which invalidates all shared locks and blocks further updates until the exclusive lock is released.

\section{State-of-the-art Snapshotting}
\label{sec:snapshotting}

As stated before, our heterogeneous processing model stands and falls with an efficient snapshot creation mechanism. Only if we are able to create them at a high frequency without penalizing the system, we get up-to-date snapshots with short version chains.
There exist different techniques to implement such a snapshotting mechanism, including \textit{physical} and \textit{virtual} techniques. While the former ones create costly physical copies of the entire memory, the latter ones lazily separate snapshots only for modified memory pages. Let us now look at the state-of-the-art techniques in detail to understand why they do not suffice our needs and why we have to introduce a completely new snapshotting mechanism in AnKerDB.

\subsection{Physical Snapshotting}
The most straightforward approach of snapshotting is \textit{physical snapshotting}, where a deep physical copy of the database is created when a snapshot is taken. 
On this physical copy, the reading queries can then run in isolation, while the modifying transactions update the original version. The granularity at which the snapshot is taken is up to the implementation. It is possible to snapshot the entire database, a table, or a set of columns.  
This way of snapshotting obviously represents the \textit{eager} way of doing it --- at the time of snapshot creation, the snapshot and the source are fully separated from each other. As a consequence, any modification to the source is not carried through to the snapshot without further handling.    

Obviously, physical snapshotting is a very straightforward approach, that is easy to use. However, its effectiveness is directly bound to the amount of data that is updated on the source. If only a portion of the data is updated, the full physical separation of the snapshot and the source is unnecessary and only adds overhead to the snapshotting cost.

\subsection{Virtual Snapshotting}

Virtual Snapshotting overcomes this problem by following the \textit{lazy} approach. The idea of virtual snapshotting is that the snapshot and the source are not separated physically when the snapshot is taken. Instead, the separation happens lazily for those memory pages, that have actually been modified. As we will see, there a multiple ways to perform this separation using virtual memory. To understand them, let us first go through some of the high-level concepts of the virtual memory subsystem of \textit{Linux} (kernel~4.8).

\subsubsection{Virtual vs Physical Memory}
\label{ssec:userperspective}

By default, the user perspective on memory is very simple --- he sees only virtual memory. 
\begin{figure}[!htb]
	\vspace{-0.2cm}
	\begin{center}
		\includegraphics[width=4in]{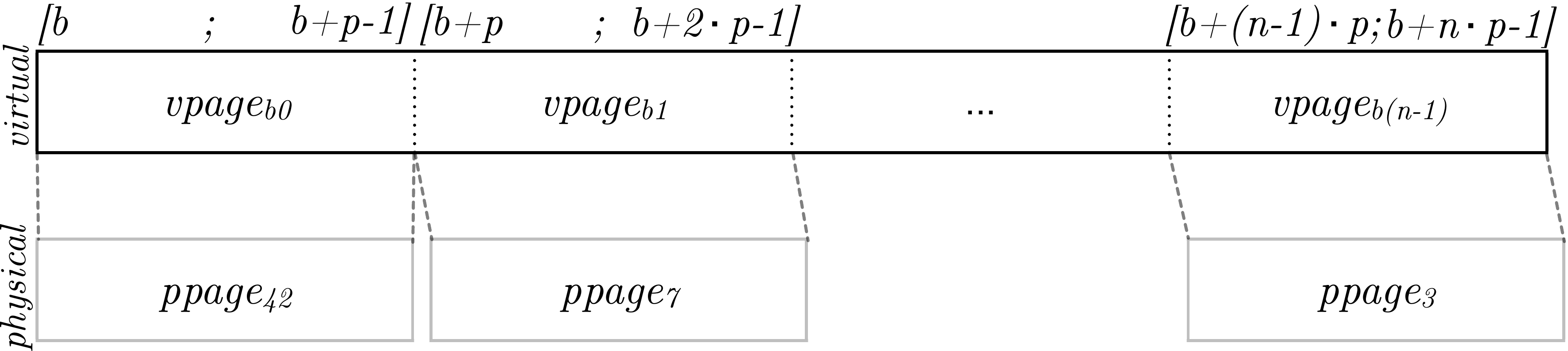}
	\end{center}
	\vspace{-0.5cm}
	\caption{\textbf{Conceptual mapping of virtual to physical memory.} The start address of the virtual memory area is denoted as~$b$ and the size of a page as~$p$. A consecutive virtual memory area of $n$~pages is transparently mapped by the operating system to $n$~scattered physical pages.}
	\label{figs:vmem}
\end{figure}

To allocate a consecutive virtual memory area~$b$ of size~$s$ the system call \smalltt{mmap} is used.

For instance, the well known general purpose memory allocator \smalltt{malloc} internally uses \smalltt{mmap} to claim large chunks of virtual memory from the operating system. The layer of physical memory is completely hidden and transparently managed by the operating system. Figure~\ref{figs:vmem} visualizes the relationship of the memory types.  
After allocating the virtual memory area, the user can start accessing the memory area, e.g. via $b[i]=42$. 
Obviously, the user perspective is fairly simple. He basically does not have to distinguish between memory types at all. In comparison, the kernel perspective is significantly more complex.

\begin{figure}[!htb]
	\vspace{-0.2cm}
	\begin{center}
		\includegraphics[width=16cm]{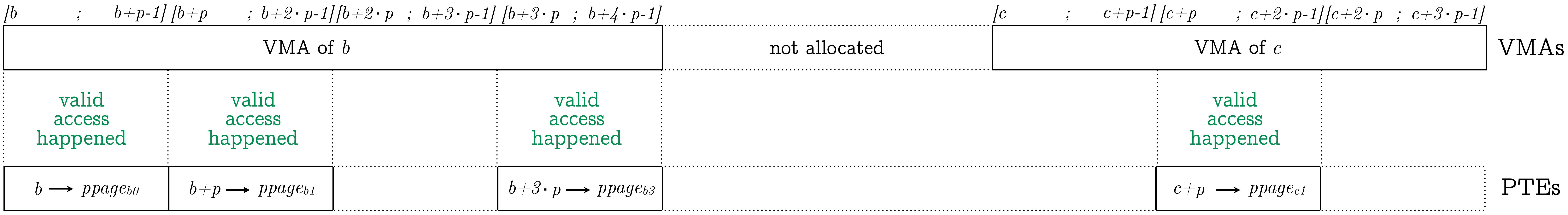}
	\end{center}
	\vspace{-0.5cm}
	\caption{\textbf{Visualization of the relationship between VMAs and PTEs}. The VMAs store the information which virtual memory areas are currently allocated alongside with all necessary meta-information. The page table contains for every accessed virtual page one entry that maps the virtual start address of that virtual page to its physical page.}
	\label{figs:vma_to_pte}
\end{figure}

First of all, the previously described call to \smalltt{mmap}, which allocates a consecutive virtual memory area, does not trigger the allocation of physical memory right away. Instead, the call only creates a so called \smalltt{vm\_area\_struct} (called \textit{VMA} in the following), that contains all relevant information to \textit{describe} this virtual memory area. For instance, it stores that the size of the area is~$s$ and that the start address is~$b$. Thus, the set of all VMAs of a process define which areas of the virtual address space are currently \textit{reserved}. Note that a single VMA can describe a memory area spanning over multiple pages. As an example, in Figure~\ref{figs:vma_to_pte} we visualize two VMAs. They describe the virtual memory areas starting at address~$b$~(spanning over four pages) respectively starting at~$c$~(spanning over three pages). In between the two memory areas is an unallocated memory area of size two pages.

Besides of the VMAs, there exists the concept of the \textit{page table} within each process.  An entry in the page table (called \textit{PTE} in the following) contains the actual mapping from a single virtual to a single physical page and is only inserted after the first access to a virtual page, based on the information stored in the corresponding VMA. The example in Figure~\ref{figs:vma_to_pte} shows the state of the page table after four accesses to four different pages happened. As we can see, we have one PTE per page in the page table.

\subsubsection{Fork-based Snapshotting}
\label{ssec:fork}

With the distinction between the different memory types and the separation of VMAs and PTEs in mind, we are now able to understand the most fundamental form of virtual snapshotting: fork-based snapshotting~\cite{hyper}. It exploits the system call \smalltt{fork}, which creates a child process of the calling parent process. This child process gets a copy of all VMAs and PTEs of the parent. In particular, this means that after a fork, the allocated virtual memory of the child and the parent share the same physical memory. Only a write\footnote{Assuming the virtual memory area written to is private (\smalltt{MAP\_PRIVATE}).} to a page of child or parent triggers the actual physical separation of that page in the two processes (called \textit{copy-on-write} or COW). 

Obviously, this concept can be exploited to implement a form of snapshotting. If the source resides in one process we simply fork it to create a snapshot. Any modification to the source in the parent process is not visible to queries running on the snapshot in the child process. As mentioned in Section~\ref{ssec:het_processing}, early versions of Hyper that implemented heterogeneous processing utilized that mechanism.

\subsubsection{Rewired Snapshotting}
\label{ssec:rewiring}

While fork-based snapshotting has the convenient advantage, that the snapshotting mechanism is handled by the operating system in a transparent fashion, it has two major disadvantages as well. First, it requires the spawning and management of several processes at a time. Second, it always snapshots all allocated memory of the process, i.e. it can not be limited to a subset of the data. Both problems can be addressed using our technique of rewiring, which we already applied on the snapshotting problem in~\cite{rewiring}. 

\begin{figure}[!htb]
	\vspace{-0.2cm}
	\begin{center}
		\includegraphics[width=4.5in]{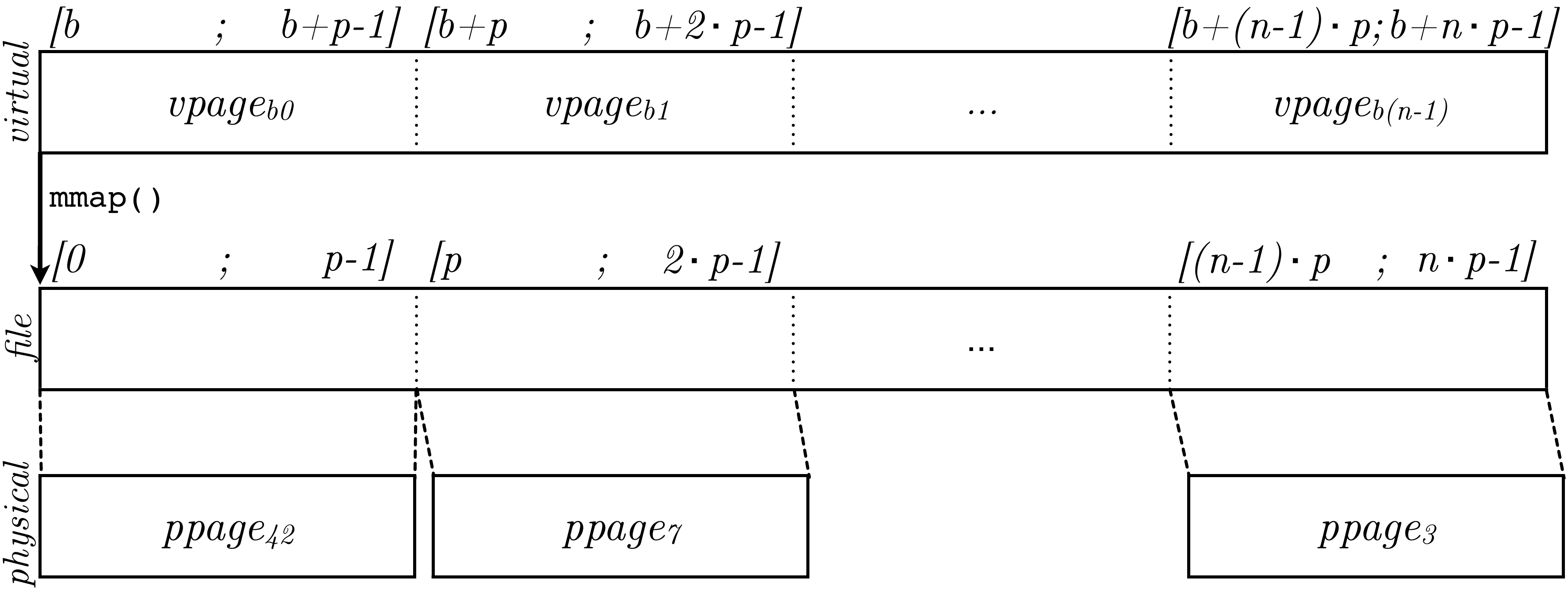}
	\end{center}
	\vspace{-0.5cm}
	\caption{In \textbf{rewiring memory}~\cite{rewiring}, we map a virtual memory area (described by a VMA) to a main memory file. A~single call to \texttt{mmap} maps multiple virtual memory pages to the main-memory file. The start address of the area is denoted as $b$. The virtual page $vpage_{bi}$ starting at address $b+i\cdot p$ is mapped to the file at offset $i\cdot p$ which in turn is backed by some physical page.}
	\label{figs:vmemtofile}
\end{figure}

To understand rewiring, let us again look at the mapping from virtual to physical memory as described in Section~\ref{ssec:userperspective}. This mapping is by default both \textit{hidden} from the user as well as \textit{static}, as the user sees only virtual memory by default. This is why we dedicated our recent work of rewiring memory~\cite{rewiring} to the reintroduction of physical memory to user space. We bring back this type of memory in the form of so called \textit{main-memory files}. As it is possible to freely map virtual memory to main-memory files using the system call~\smalltt{mmap} and main-memory files are internally backed by physical memory, we have established a transitive mapping from virtual to physical memory. This mapping can be updated at any time. Figure~\ref{figs:vmemtofile} shows the concept. This means using rewiring memory, we established a mapping that is both \textit{visible} and \textit{modifiable} in user space.

In rewired snapshotting, we utilize this modifiable mapping. Let us assume we have a virtual memory area~$b$ as shown in Figure~\ref{figs:vmemtofile}, on which we want to create a snapshot. To snapshot, we simply allocate a new virtual memory area~$c$ and rewire it to the file, which represents our physical memory, in the same way as~$b$. Consequently, $b$ and $c$ share the same physical pages. If now a write to a page of $b$ is happening, the separation of snapshot and original version must be performed manually on that page, before the write can be carried out. In the first place, we have to detect the write. 
After detection, we claim an unused page from the file (which serves as our pool for free pages), copy the content of the page over, perform the write, and rewire~$b$ to map to the new page.

By this, we are able to mimic the behavior of \smalltt{fork} while staying within a single process. Further, we can offer the flexibility of snapshotting only a fraction of the data. However, rebuilding the mapping can also be quite expensive as we will see in the following.

\subsection{Reevaluating the State-of-the-Art}
\label{sec:limitations}

As we have discussed the different state-of-the-art methods of physical and virtual snapshotting that are present, let us now try to understand their individual strengths and limitations. This analysis will point us directly to the requirements we have on our custom system call, that we will use in AnKerDB to power snapshotting.

In the experiment we are going to conduct in Section~\ref{ssec:creating_snapshot}, we evaluate the time to \textit{create a snapshot} in the sense of a establishing a separate view on the data. While for physical snapshotting, this means creating a deep physical copy of the data, for virtual snapshotting, it does not trigger any physical copy of the data. Still, virtual snapshotting has to perform a certain amount of work as we will see.  
We will perform the experiment as a stand-alone micro-benchmark to focus entirely on the snapshotting costs and to avoid interference with other components, that are present in a complex transactional processing system like AnKerDB. We use a table with $n=50$ columns that is stored in a columnar fashion, where each column has a size of $200$MB.

The question remains which page size to use. To make snapshotting as efficient as possible, we want to back our memory with pages \textit{as small as available}. This ensures that the overhead of copy-on-write on the level of page granularity is minimal.
Consider the case where our $200$MB column is either backed by $100$~huge pages or $51{,}200$~small pages. In the former case, $100$~uniformly distributed writes would cause a COW of the entire column ($200$MB) in the worst case, resulting in a full physical separation of the snapshotted column and the base column.
In the latter case, $100$~writes would trigger COW of only $100$~small pages ($400$KB), physically separating only $0.2\%$ of the snapshotted column from the base column.

\subsubsection{System Setup}
We perform all of the following experimental evaluations on a server consisting of two quad-core Intel Xeon E5-2407 running at a clock frequency of $2.2$~GHz. The CPU does neither support hyper-threading nor turbo mode. The sizes of the L1 and L2 caches are $32$KB and $256$KB, respectively, whereas the shared L3 cache has a size of $10$MB. The processor can cache 64 entries in the fast first-level data-TLB for virtual to physical 4KB page address translations. In a slower second-level TLB, 512 translations can be stored. For $2$MB huge pages, the TLB can cache 32 translations in L1 dTLB. In total, the system is equipped with $48$GB of main memory, divided into two NUMA regions. For all experiments, we deactivate one CPU and the attached NUMA region to stay local on one socket. The operating system is a 64-bit version of Debian~8.16 with Custom Linux kernel version~4.8.17. The codebase is written in C++ and compiled using g++~6.3.0 with optimization level~O3.

\subsubsection{Creating a Snapshot}
\label{ssec:creating_snapshot}
To simulate snapshotting on a subset of the data, we create a snapshot on the first $p$~columns of the table~$T$. 
Let us precisely define how the individual snapshotting techniques behave in this situation:

\textbf{(a)~Physical}:~to create a snapshot of $p$~columns of table~$T$, we allocate a fresh virtual memory area~$S$ of size $p \cdot l$~pages. Then, we copy the content of $p$~columns of $T$ into~$S$ using \smalltt{memcpy}. $S$~represents the snapshot. 

\textbf{(b)~Fork-based}:~to create a snapshot of $p$~columns of table~$T$, we create a copy of the process containing table~$T$ using the system call~\smalltt{fork}. Independent of~$p$, this snapshots the entire table. The first $p$~columns of table~$T'$ contained in the forked process represent the snapshot. The virtual memory areas representing $T$ and $T'$ are declared as \textit{private}, such that writes to one area are not propagated to the other area.

\textbf{(c)~Rewiring}:~to create a snapshot of $p$~columns of table~$T$, we first have to inspect by how many VMAs each column is actually described. As a VMA describes the common properties of a consecutive virtual memory region, 
it is possible that a column is described by only a single VMA (best case), by one VMA per page (worst case), or anything in between. The more writes happened to a column and the more copy-on-writes were performed, the more VMAs a column is backed by. Eventually, every page is described by its individual VMA.

To create the snapshot, we first allocate a fresh virtual memory area~$S$ of size $p \cdot l$~pages. 
For each VMA that is backing a portion of the $p$~columns in $T$, we now rewire the corresponding portion of~$S$ to the same file offset.
Additionally, we use the system call~\smalltt{mprotect} to set the protection of $S$ to read-only. This is necessary to detect the first write to a page to perform a manual copy-on-write. $S$ represents the snapshot.

Table~\ref{table:snapshot_creation} shows the results.
We vary~$p$, the number of columns to snapshot, from $1$~column ($2\%$ of the table) over $25$~columns ($50\%$ of the table) to $50$~columns ($100\%$ of the table) and show the runtime in ms to create the snapshot. For rewiring we vary the pages that have been modified (by writing the first $8$B of the page) before the snapshot is taken, as it influences the runtime. We test the case where no write has happened and each column is backed by a single VMA. Further, we measure the snapshotting cost after $500$~pages, $5{,}000$~pages, and $50{,}000$ pages have been modified. These number of writes lead to $995$, $9483$, and $51177$ number of VMAs backing a column respectively.

First of all, we can see that physical snapshotting is quite expensive, as it creates a deep copy of the columns already at snapshot creation time. As expected, we can observe a linearly increasing cost with the number of columns to snapshot. In contrast to that, fork-based snapshotting is independent from the number of requested columns, as it snapshots the entire process with the entire table in any case. When snapshotting $50\%$ of the table, fork-based snapshotting is over an order of magnitude faster than physical snapshotting, as it duplicates solely the virtual memory, consisting of the VMAs and the page table.
The runtime of rewiring is highly influenced by the number written pages respectively the number of VMAs per column. The more VMAs we have to touch to create the snapshot, the higher the runtime. If we have as many VMAs as pages (which is essentially the case after $50{,}000$~writes), the performance of rewiring pretty much equals the one of physical snapshotting. However, we can also see rewiring is significantly faster than the remaining methods, if less VMAs need to be copied. For instance, after $500$~writes, rewiring is around two orders or magnitude faster for a single column and still almost factor two faster for snapshotting the entire table. 

\begin{table}[!htb]
\vspace{-0.2cm}
	\setlength{\tabcolsep}{3pt}
	\begin{center}
		\begin{tabular}{| L{2.8cm} | L{3cm} || R{3cm} | R{3cm} | R{3cm} |}
			\hline
			\textbf{Method} & \textbf{Pages Modified per Col} & \textbf{1 Col} [ms] & \textbf{25 Col} [ms] & \textbf{50 Col} [ms]\\
			\hline
			\hline
			Physical & -- & $108.09$ & $2693.69$ & $5382.87$\\
			\hline
			Fork-based & -- & $108.28$ & $108.28$ & $108.28$ \\
			\hline
			Rewiring & $0$ & $0.02$ & $0.39$ & $7.72$\\
			\hline
			Rewiring & $500$ & $1.22$ & $30.90$ & $61.87$\\
			\hline
			Rewiring & $5{,}000$ & $14.17$ & $352.15$ & $712.96$\\
			\hline
			Rewiring & $50{,}000$ & $169.28$ & $4210.17$ & $8459.67$\\
			\hline
		\end{tabular}
		\vspace{-0.1cm}
		\caption{\textbf{Creating a snapshot} using the state-of-the-art techniques. We vary the number of columns on which the snapshot is taken. For rewiring, the number of modified pages has a drastic impact on the runtime. Thus, we show the snapshotting cost after $0$, $500$, $5{,}000$, and $50{,}000$ pages were modified per column.
		}
		\label{table:snapshot_creation}
	\end{center}
\end{table}

\subsubsection{Summary of Limitations}

Obviously, the performance of rewiring for snapshot creation is highly influenced by the number of VMAs per column. For every VMA, a separate \smalltt{mmap} call must be carried out -- a significant cost if the number of VMAs is large. Unfortunately, when using rewiring, an increase in the amount of VMAs over time is not avoidable. 

Still, we believe in rewiring for efficient snapshotting. It simply can not show its full potential. 
If we carefully inspect the description of rewired snapshotting in Section~\ref{ssec:creating_snapshot} again, we can observe that we actually implemented a \textit{workaround} of the limitations of the OS. We manually rewire the virtual memory areas described by the individual VMAs to create a snapshot --- because there is no way to simply copy a virtual memory area. We perform another pass over the VMAs to set the protection using the system call \smalltt{mprotect} to read-only --- instead of setting it directly when copying the virtual memory area.  

Obviously, we hit the limits of the vanilla kernel. Therefore, in the following Section, we will propose a custom system call that tackles these limitations --- leading to a much more straight-forward and efficient implementation of virtual snapshotting, which we will finally use in AnKerDB.

\section{System Call vm\_snapshot}
\label{sec:rewiringplusplus}

As we have seen the limitations of the state-of-the-art kernel in the previous section, let us discuss how we can overcome them. 

\subsection{Snapshotting Virtual Memory}
\label{ssec:snapshotting}

In our implementation of rewired snapshotting, we have experienced the need to directly snapshot virtual memory areas. By default, this is not supported by the kernel. As a workaround, we had to rewire a fresh virtual memory area in the same way as the source area. This is a very costly process as it involves repetitive calls to \smalltt{mmap}. 

\subsubsection{Semantics}
\label{sssec:semantics}

To solve this problem, we have to introduce a new system call, that will be the core component of our snapshotting mechanism. Before doing this, let us precisely define what \textit{snapshotting a virtual memory area} means in this context. Assuming we have a mapping from $n$~virtual to $n$~physical pages as shown in Figure~\ref{figs:vmem} of Section~\ref{ssec:userperspective}, starting at virtual address~$b$. As we can see, the first virtual page covering the virtual address space~$[b;b+p-1]$ ($vpage_{b0}$) is mapped to the physical page~$ppage_{42}$. The second virtual page covering virtual address space~$[b+p;b+2 \cdot p-1]$ ($vpage_{b1}$) is mapped to another physical page~$ppage_{7}$ and so on. Now, we want to create a new virtual memory area starting at a new virtual address (let us call it~$c$) that maps to the \textit{same} physical pages. Thus, the virtual page covering $[c;c+p-1]$ should map to $ppage_{42}$, the virtual page covering $[c+p;c+2 \cdot p-1]$ should map to $ppage_{7}$ and so on.
We define the following system call to encapsulate the described semantics:

\noindent
\begin{minipage}{\linewidth}
\begin{small}
\begin{lstlisting}
void* vm_snapshot(void* src_addr, size_t length);
\end{lstlisting}
\label{listings:vmsnapshot}
\end{small}
\end{minipage}

\noindent This system call takes the \smalltt{src\_addr} of the virtual memory area to snapshot and the \smalltt{length} of the area to copy in bytes. Both \smalltt{src\_addr} and \smalltt{length} must be page aligned. It returns the address of a new virtual memory area of size \smalltt{length}, that is a snapshot of the virtual memory area starting at \smalltt{src\_addr}. The new memory area uses the same update semantics as the source memory area, i.e. if the virtual memory area at \smalltt{src\_addr} has been declared using \smalltt{MAP\_PRIVATE | MAP\_ANONYMOUS}, the new memory area is declared in the same way.

\subsubsection{Implementation}
Implementing a system call that modifies the virtual memory subsystem of Linux is a delicate challenge. In the following, we will provide a high-level description of the system call behavior. For the interested reader, we provide a more detailed discussion in Appendix~\ref{appendix:syscalls} respectively the actual source code, that will be released along with this paper.
On a high level, \smalltt{vm\_snapshot} internally performs the following steps: (1)~Identify all VMAs that describe the virtual memory area $[\smalltt{src\_addr},\smalltt{src\_addr}+\smalltt{length})$. (2)~Reserve a new virtual memory area of size~\smalltt{length} starting at virtual address~\smalltt{dst\_addr}. (3)~Copy all of the previously identified VMAs and update them to describe the corresponding portions of virtual memory in $[\smalltt{dst\_addr},\smalltt{dst\_addr}+\smalltt{length})$. (4) For each VMA which describes a private mapping (which is the standard case in AnKerDB), additionally copy all existing PTEs and update them to map the corresponding virtual pages in $[\smalltt{dst\_addr},\smalltt{dst\_addr}+\smalltt{length})$.

This system call~\smalltt{vm\_snapshot} will form the core component of creating snapshots on columns in AnKerDB. It is the call that we use in Figure~\ref{figs:ankerdb_concept} in Step~\circled{4} and Step~\circled{7}. 

\subsubsection{Snapshotting to Existing Virtual Memory Area}

So far, our system call~\smalltt{vm\_snapshot} returns the snapshot in form of the start address to a \textit{new} virtual memory area. However, there might be situations in which we would like to realize the snapshot in an \textit{existing} virtual memory area. Therefore, we extend our system call by adding a third argument~\smalltt{dst\_addr}:

\noindent
\begin{minipage}{\linewidth}
\begin{small}
\begin{lstlisting}
void* vm_snapshot(void*  dst_addr,
                  void*  src_addr, 
                  size_t length);
\end{lstlisting}
\label{listings:vmsnapshot}
\end{small}
\end{minipage}

\noindent If~\smalltt{dst\_addr} is \smalltt{NULL}, \smalltt{vm\_snapshot} provides the semantics described in Section~\ref{sssec:semantics}, returning the address of a new virtual memory. If \smalltt{dst\_addr} is a valid address, the snapshot of $[\smalltt{src\_addr},\smalltt{src\_addr}+\smalltt{length})$ is created in $[\smalltt{dst\_addr},\smalltt{dst\_addr}+\smalltt{length})$. If $[\smalltt{dst\_addr},\smalltt{dst\_addr}+\smalltt{length})$ is not (entirely) allocated, the call fails.

This extension to \smalltt{vm\_snapshot} allows us to reuse previously allocated virtual memory areas. For instance, when replacing an outdated snapshot of a column with a fresh one, we can simply ``recycle" its allocated virtual memory area.

\subsubsection{Evaluation}

Let us now see how our custom system call~\smalltt{vm\_snapshot} performs in comparison with its direct competitor rewiring. We excluded the baselines of physical snapshotting and fork-based snapshotting, as they are already out of consideration for AnKerDB due to high cost and low flexibility. We first look again at the snapshot creation time for a single column of $200$MB. The previous experiment presented in Table~\ref{table:snapshot_creation} showed that rewiring is highly influenced by the number of VMAs that are backing the column to snapshot. To analyze this behavior in comparison with \smalltt{vm\_snapshot}, we run the following experiment: for each of the $51{,}200$ pages of the column, we perform exactly one write to the first $8$B of the page. In the case of rewiring, this write triggers the COW of the touched page and thus creates a separate VMA describing it. After each and every write, we create a new snapshot of the column and report the time of snapshot creation. 

\begin{figure}[!htb]
\centering
\subfloat[\textbf{Comparison of snapshot creation times}. The time to take a snapshot is shown on the left $y$-axis for rewiring respectively \smalltt{vm\_snapshot}. To enhance the visualization, we also show a zoom-on of the origin.]{
\includegraphics[width=8cm, trim={0 0 0 0}, clip]{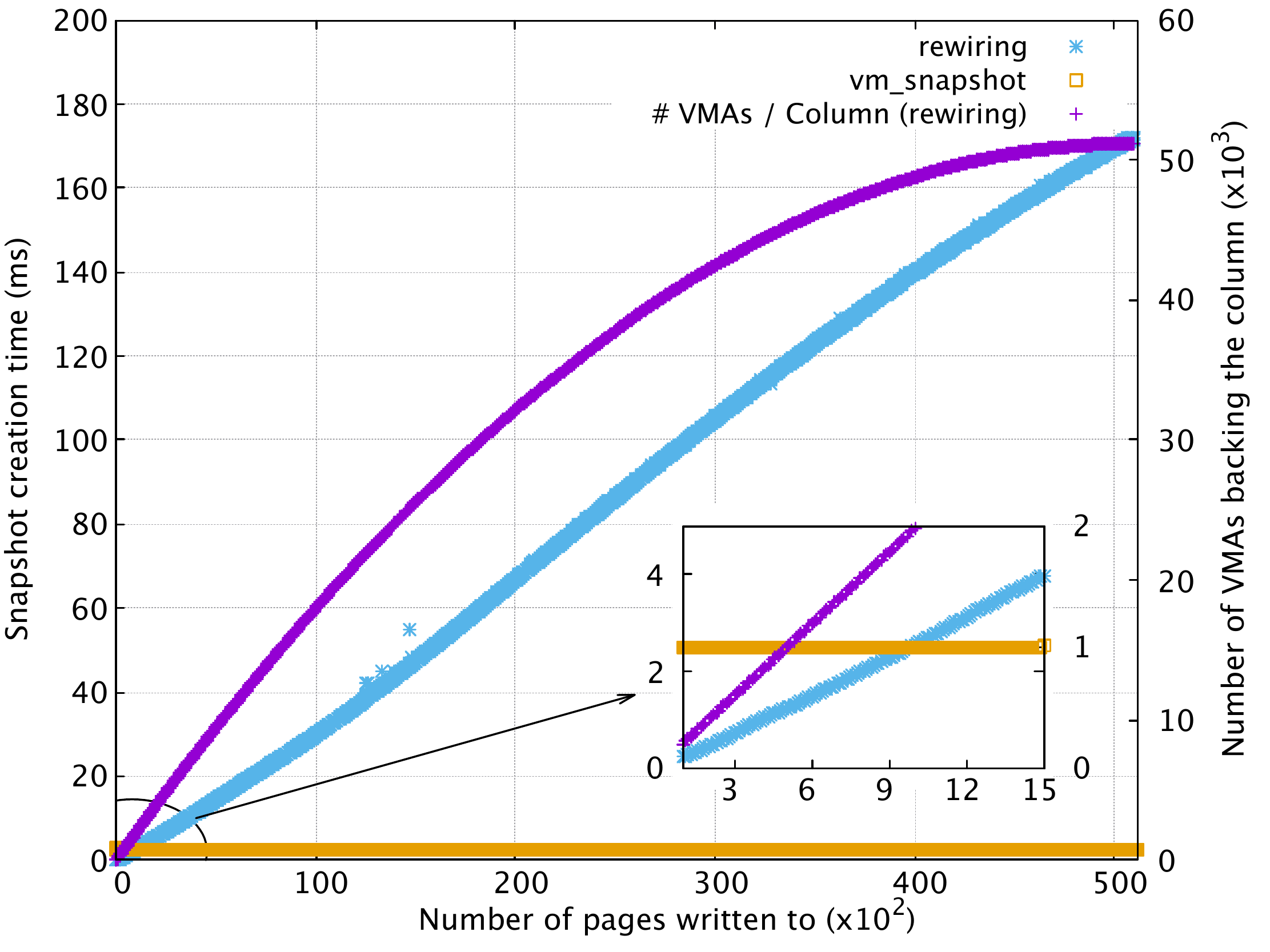}
\label{figs:snapshot_creation_time}
}\\
\subfloat[\textbf{Comparison of writes} to the snapshotted column. On the left $y$-axis, the time to perform a write of $8$B is shown.]{
\includegraphics[width=8cm, trim={0 0 0 0}, clip]{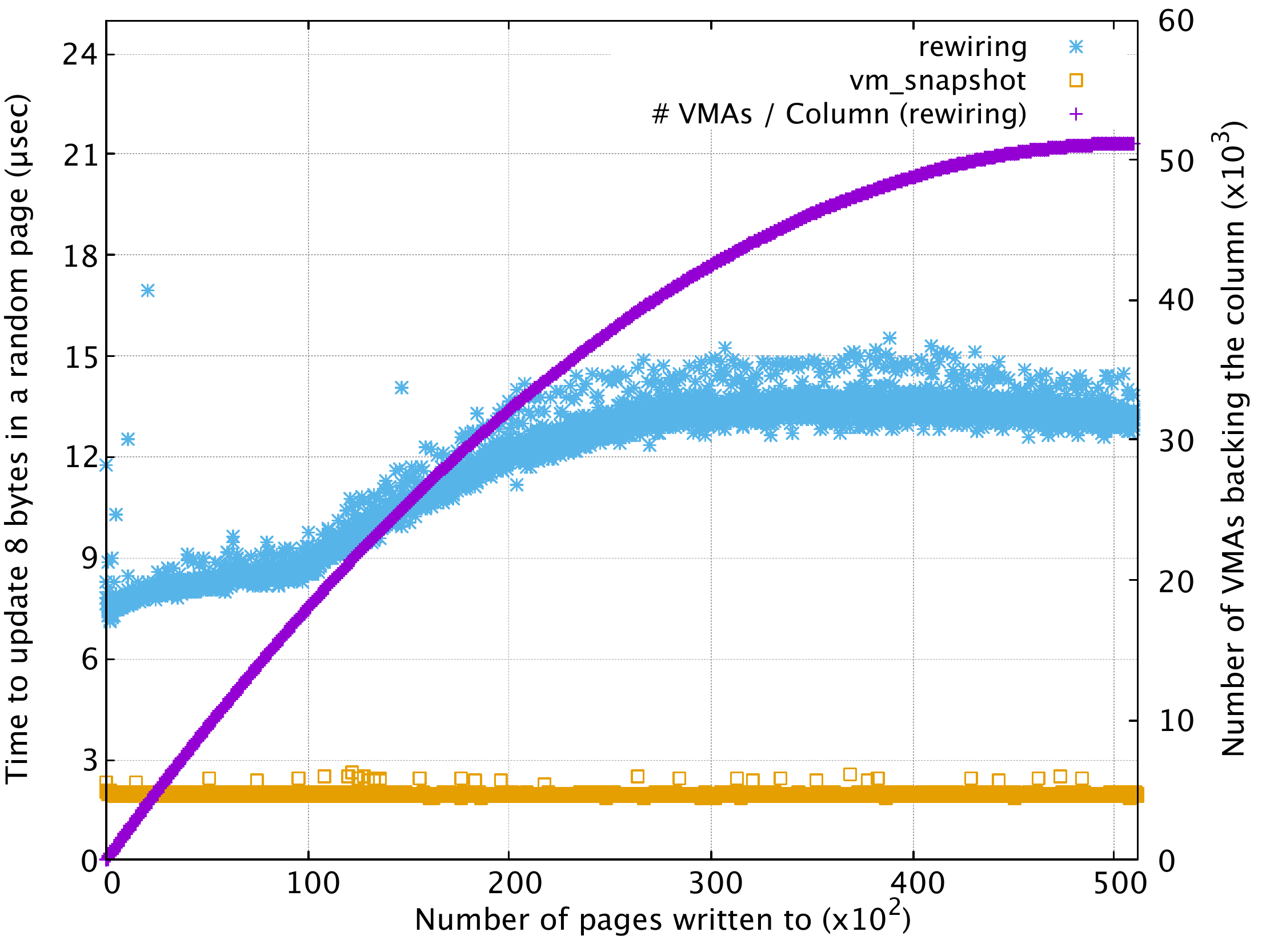}
\label{figs:write_cost}
}
\caption{\textbf{Comparison of vm\_snapshot and rewiring} in terms of snapshot creation cost and write cost. After every write to a page, a new snapshot is taken. Figure~\ref{figs:snapshot_creation_time} shows the snapshot creation times and Figure~\ref{figs:write_cost} shows the cost of the writes on the left $y$-axis. Alongside, we show the number of VMAs per column for rewiring on the right $y$-axis.}
\label{figs:vm_snapshot_comp}
\end{figure}

Let us look at the results in Figure~\ref{figs:snapshot_creation_time}. As predicted, the snapshot creation cost of rewiring is highly influenced by the number of VMAs that is increasing with every modified page. To visualize this correlation, we plot the number of VMAs per column for rewiring alongside with the snapshot creation time. In contrast to rewiring, our system call~\smalltt{vm\_snapshot} shows both a very stable and low runtime over the entire sequence of writes. After only around $1000$~writes have happened (see zoom-in of Figure~\ref{figs:snapshot_creation_time}), the snapshotting cost of \smalltt{vm\_snapshot} already becomes lower than than the one of rewiring. After all $51{,}200$ writes have been carried out, \smalltt{vm\_snapshot} is $68$x faster than rewiring. This shows the tremendous effect of avoiding repetitive system calls to~\smalltt{mmap}.  

However, snapshot creation time is not the sole cost to optimize for. We should also look at the actual cost of writing the virtual memory. In the case of rewiring, the triggered COW is handled manually by copying the page content to an unused page and rewiring that page into the column. In the case of \smalltt{vm\_snapshot}, which works on anonymous memory and relies on the COW mechanism of the operating system, no manual handling is necessary. This becomes visible in the runtime shown in Figure~\ref{figs:write_cost}. Obviously, writing a page of the column snapshotted by \smalltt{vm\_snapshot} is up to $6$x~faster than writing to one created by rewiring. The reason for this is that the entire COW is handled by the operating system. No protection must be set manually, no signal handler is necessary to detect the write to a page. 

\section{Experimental Evaluation}
\label{sec:experimental_evaluation}

After the description of AnKerDB's system design and the introduction of our custom system call \smalltt{vm\_snapshot} to efficiently snapshot virtual memory areas, let us now start with the experimental evaluation of the actual system. As AnKerDB relies on a heterogeneous processing model, we want to test it against the homogeneous counterparts. AnKerDB is designed in a way to also support homogeneous processing via configuration by disabling snapshotting. 

\subsection{AnKerDB Configurations}
\label{ssec:configurations}

Let us look at the different configurations we are going to evaluate:
\begin{enumerate}

\item \label{hom_fs} \textbf{Homogeneous processing, full serializability}. We configure AnKerDB such that \textit{no snapshots} are taken at all. Thus, there is only the OLTP component with the most recent representation of the database. Both OLTP and OLAP transactions run in the OLTP component under \textit{full serializability} guarantees. As in this setup version chains build up that are not discarded automatically with snapshots, a garbage collection mechanism is necessary. We use a thread that makes a pass over the version chains every second and deletes all versions that are older than the oldest transaction in the system.

\item \label{hom_si} \textbf{Homogeneous processing, snapshot isolation}. As in~(\ref{hom_fs}.), \textit{no snapshots} are taken. There is only the OLTP component with the most recent representation of the database. Both OLTP and OLAP transactions run in the OLTP component under \textit{snapshot isolation} guarantees and thus, the read set validation is not performed. The same garbage collection mechanism as in (\ref{hom_fs}.) is applied.  

\item \label{het} \textbf{Heterogeneous processing, full serializability}. The OLTP transactions run in the OLTP component and the OLAP transactions run in the OLAP component. The creation of snapshots works as described in Section~\ref{ssec:sync}: after a certain amount of commits to the database has been registered ($10{,}000$ in the upcoming experiments), the system sets a snapshot timestamp, that will mark the time of the snapshot to create. The very next access which a column receives will now trigger the actual snapshot creation using our system call. By this, we are able to generate snapshots that are consistent with respect to a single point in time but that are also created in a lazy fashion based on the actual access pattern. 

\end{enumerate}

\subsection{Experimental Setup}
\label{ssec:experimental_setup}

To evaluate the system under complex transactions, we define the following mixture of OLTP and OLAP transactions:

On the side of OLAP, we form transactions based on queries of the TPC-H~\cite{tpch} benchmark. Precisely, we pick the single table queries~Q1 and Q6 (\smalltt{LINEITEM}) and Q4~(\smalltt{ORDERS}) as well as the two-table query~Q17 (joining~\smalltt{LINEITEM} and \smalltt{PART}) as good representatives. For each fired OLAP transaction, we pick the configuration parameters of the query randomly within the bounds given in the TPC-H specification. Additionally, for each table (\smalltt{LINEITEM}, \smalltt{ORDERS}, and \smalltt{PART}) we add a simple scan transaction that runs over the respective table. Thus, in total, we have $7$~OLAP transactions.

\begin{figure}[!htb]
	\vspace{-0.2cm}
	\begin{center}
		\includegraphics[width=11cm, trim={0.2cm 4cm 15.7cm 0}, clip]{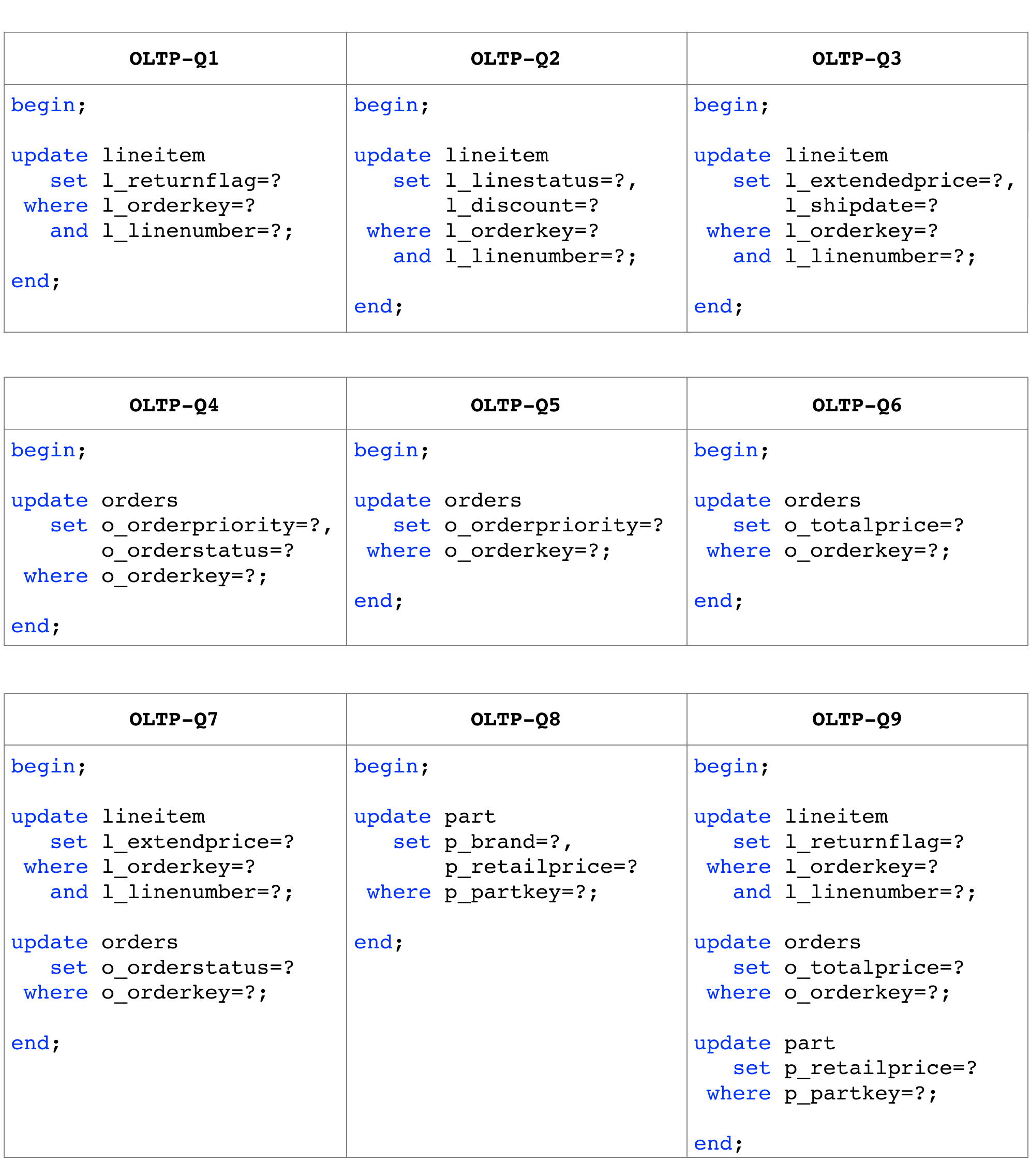}
	\end{center}
	\vspace{-0.5cm}
	\caption{The \textbf{9 OLTP transactions} we introduce and apply in the experimental evaluation. The question marks denote the parameters.}
	\label{figs:complex_oltp_transactions}
	\vspace{-0.2cm}
\end{figure}

On the OLTP side, we introduce $9$~artificial transactions. Instead of relying on queries given by a transactional benchmark (like \mbox{TPC-C}), we decided to introduce hand-tailored transactions. The reason for this is, that the transactions specified in benchmarks are typically quite large and thus very hard to control and to configure. As a consequence, results that are based on these transactions are even harder to interpret. However, since our system design is focused on improving on the OLAP throughput, we need controllable OLTP transactions to precisely adjust the OLTP load on the system. This allows us to carefully inspect the impact on the OLAP side. In this regard, we introduce the set of OLTP transactions as depicted in~Figure~\ref{figs:complex_oltp_transactions}. The question marks denote the transactions parameters, that we set when firing the transactions. For the \smalltt{VARCHAR} attributes \smalltt{l\_returnflag}, \smalltt{l\_linestatus}, \smalltt{o\_orderpriority}, and \smalltt{p\_brand}, we pick an existing value from the column in a uniform and random fashion. For the \smalltt{DOUBLE} attributes \smalltt{l\_discount}, \smalltt{l\_extendedprice}, \smalltt{o\_totalprice}, and \smalltt{p\_retailprice}, which we update in the transactions, we take the current value at the selected row and increment it by $\pm x\%$ with $x\in\{1...10\}$. In the same fashion, the \smalltt{DATE} attribute \smalltt{l\_shipdate} is updated by incrementing the current value by $\pm x$ days, with~$x\in\{1...10\}$.

\subsection{OLAP Transaction Latency}
\label{ssec:latency}

\begin{figure}[!htb]
	\vspace{-0.2cm}
	\begin{center}
		\includegraphics[width=15cm, trim={0 0 0 0}, clip]{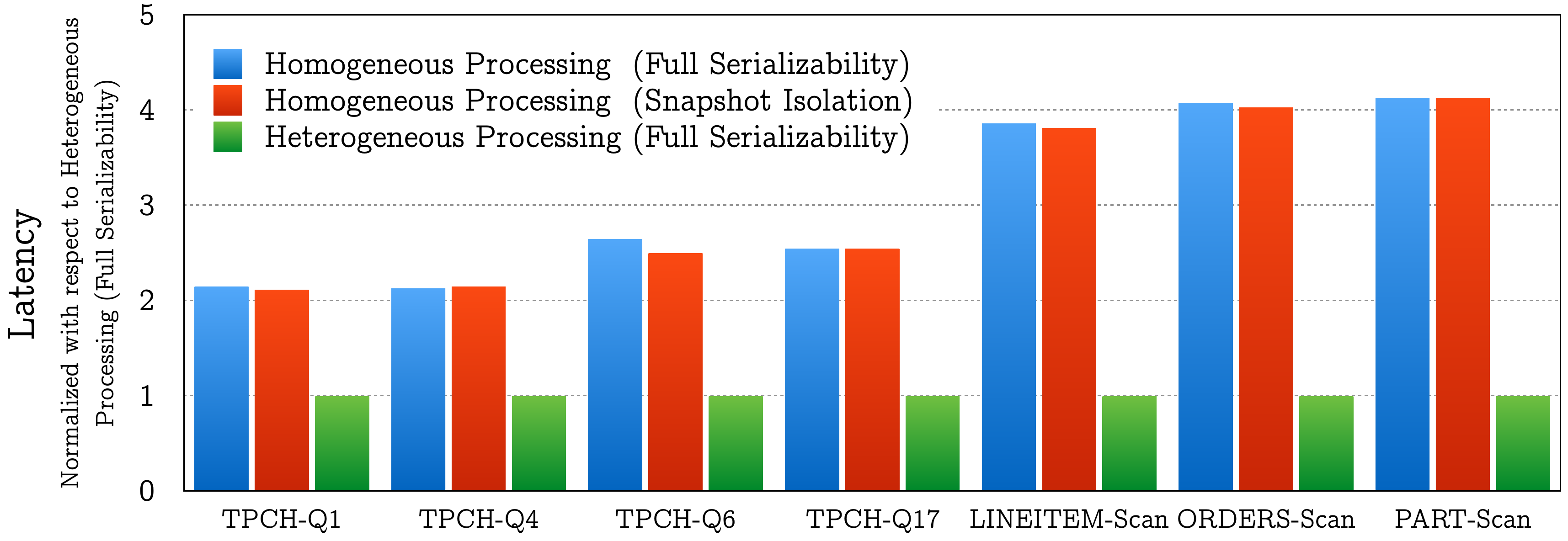}
	\end{center}
	\vspace{-0.5cm}
	\caption{\textbf{Latency of OLAP Transactions}. While a stream of OLTP transactions pressurizes $7$~out of $8$~threads of the system, we fire the respective OLAP transaction in total five times and take the average latency. We show the latency of homogeneous processing normalized with respect to heterogeneous processing.}
	\label{figs:heterogeneous_vs_homogeneous_olap_8T}
\end{figure}

Let us start the evaluation by looking at the \textit{transaction latency} in Figure~\ref{figs:heterogeneous_vs_homogeneous_olap_8T}. Precisely, we want to identify the response time for an individual OLAP transaction if the system is under load.

To measure the latency of an OLAP transaction, we pressurize the system by executing $500{,}000$ OLTP transactions picked randomly from the set of transactions described in Figure~\ref{figs:complex_oltp_transactions}. These OLTP transactions are worked by $7$~threads while the $8$th thread answers the OLAP transaction, for which we want to measure the latency. As described before, every $10{,}000$~commits a snapshot creation is triggered. To get stable results, we fire the OLAP transaction five times in total, measure the latency for each and use the average. We perform this experiment for the two homogeneous baseline configurations as well as for our heterogeneous setup and report the latency of the baselines normalized with respect to our heterogeneous approach. 

In Figure~\ref{figs:heterogeneous_vs_homogeneous_olap_8T}, we can see that for all OLAP transactions, heterogeneous processing achieves a significantly lower latency than the homogeneous baseline configurations. Our approach is around $2$x to $4$x faster depending on the tested OLAP transaction. The reason for this is that under heterogeneous processing, the OLAP transactions run entirely on snapshots in the OLAP component. While the OLAP transaction is running, the OLTP transactions perform the updates in complete isolation inside the OLTP component. In contrast to that, in the case of homogeneous processing, the updates push new versions, which are possibly relevant for the OLAP transactions, into the version chain. This results in expensive repetitive checks of timestamps at access time, heavily slowing down the scans performed by the OLAP transactions. 
In contrast to that, the OLAP transaction running on the snapshot can scan the column entirely in-place in a tight loop, without considering the version chains at all. 

\subsection{Transaction Throughput}
\label{ssec:throughput}

\begin{figure}[!htb]
	\vspace{-0.2cm}
	\begin{center}
		\includegraphics[width=10cm, trim={0 0 0 0}, clip]{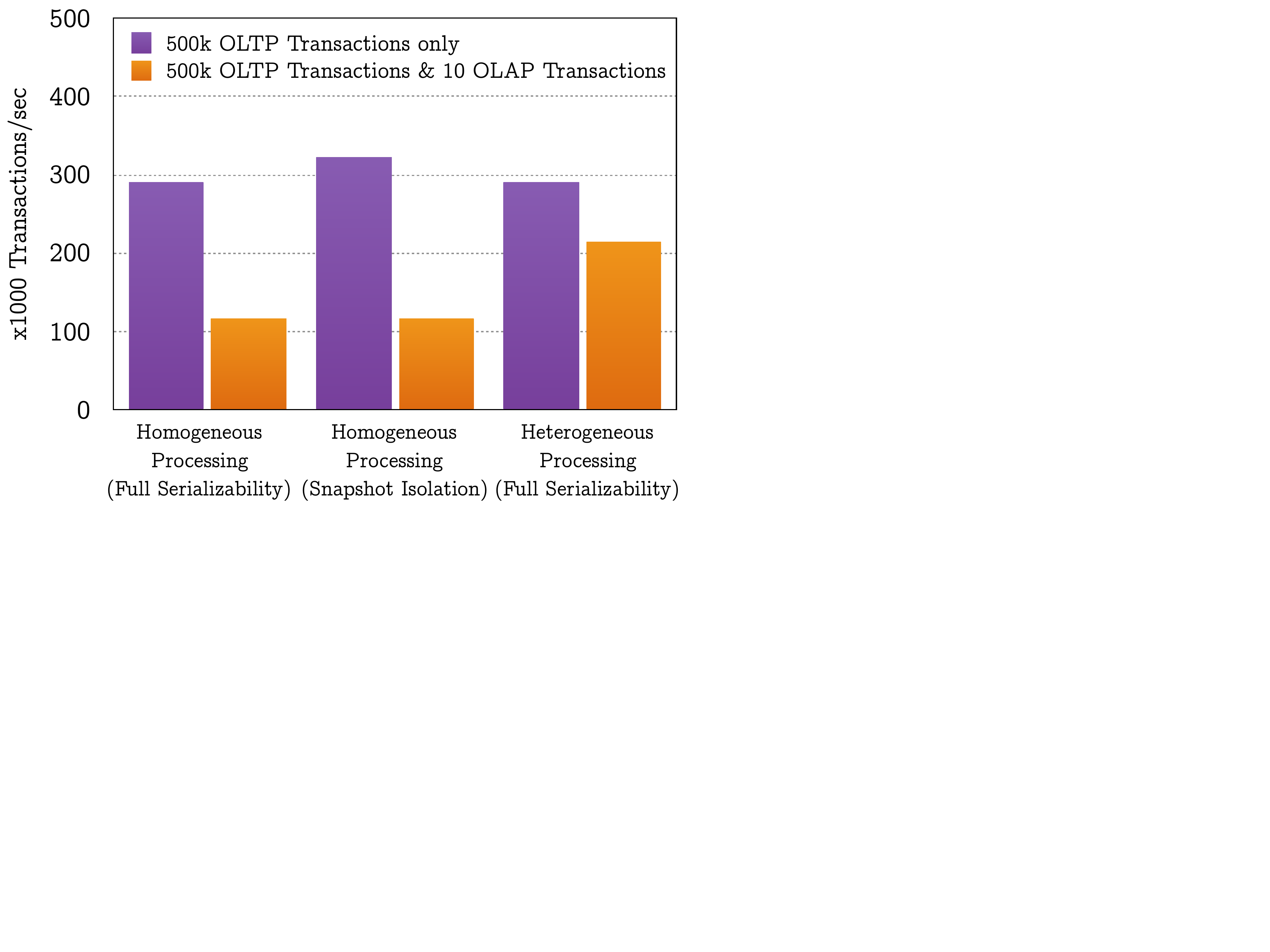}
	\end{center}
	\vspace{-0.5cm}
	\caption{\textbf{Throughput of Transaction Processing} in transactions per second. In the first experiment (violet bars), we fire only OLTP transactions. In the second experiment (orange bars), we additionally fire $10$~OLAP transactions.}
	\label{figs:throughput}
\end{figure}

Let us now look at a traditional property in estimating the quality of a transaction processing system: the throughput at which a batch of transactions can be answered from end to end. To find out, we perform the two experiments that are depicted in Figure~\ref{figs:throughput}. 

In the first experiment, presented by the violet bars, we fire $500{,}000$~OLTP transactions and process them with all~$8$ threads of our system. As before in the latency experiment, every $10{,}000$~commits we create a fresh snapshot. As expected, the throughput under snapshot isolation is the highest of all configurations, as no commit phase validation must be performed. More interesting for us in the fact that the OLTP throughput under the heterogeneous processing model equals the one under homogeneous processing. This essentially means that our heterogeneous design including snapshotting, aiming at improving OLAP processing, does \textit{not} negatively influence OLTP throughput. 

In the second experiment, depicted in the orange bars, we want to evaluate the throughput under a mixed workload. Additionally to firing $500{,}000$~OLTP transactions, we also fire $10$~OLAP transactions picked from the set of $4$~TPC-H transactions and the $3$~full table scans that we specified before in Section~\ref{ssec:experimental_setup}.  
As we can see, the mixed workload is where the heterogeneous design shines. Heterogeneous processing achieves a throughput that is almost by a factor~of~$2$ higher than the baselines. This shows the importance of separating OLAP from OLTP transactions in different processing components.  

\subsection{MVCC Scan Performance}
\label{ssec:scan_performance}

\begin{figure}[!htb]
	\vspace{-0.2cm}
	\begin{center}
		\includegraphics[width=\columnwidth, trim={0 0 0 0}, clip]{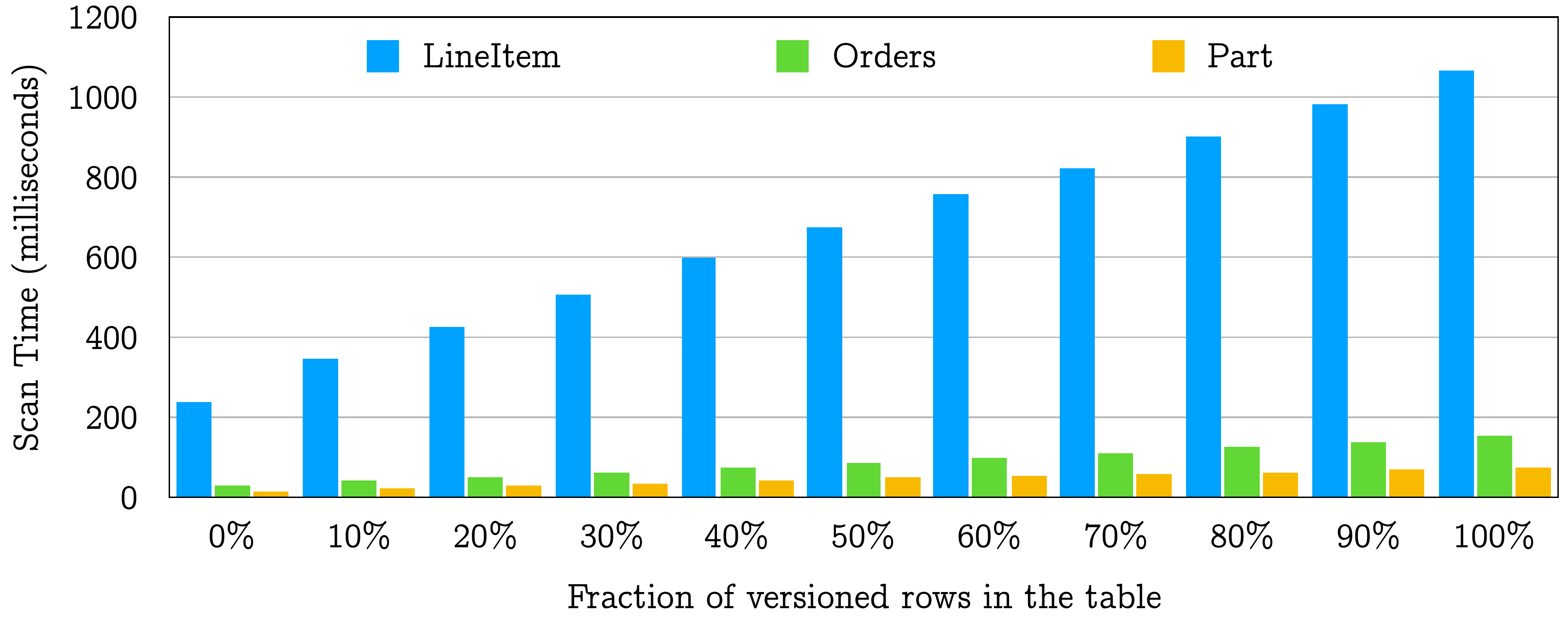}
	\end{center}
	\vspace{-0.5cm}
	\caption{\textbf{Runtime of scanning versioned tables}. We vary the amount of versioned rows and perform a full scan.}
	\label{figs:scan_over_versioned}
\end{figure}

In the previous section, we have seen that for mixed workloads, the throughput is significantly higher under our heterogeneous design. This is largely caused by the fact that OLAP transactions can simply scan the snapshotted column(s) in a tight loop instead of inspecting timestamps and traversing version chains. To investigate the problems connected with running OLAP transactions over versioned columns, we perform the experiment shown in Figure~\ref{figs:scan_over_versioned}, which resembles executing mixed workloads under homogeneous processing.

\noindent In this experiment, we vary the number of rows in the table that are versioned for \smalltt{LINEITEM}, \smalltt{ORDERS}, and \smalltt{PART} and measure the time it takes to perform a full scan of the table. The versioned rows are uniformly distributed across the table. To improve scan performance in the presence of versioned rows, we apply an optimization technique introduced by HyPer~\cite{hypermvcc}: for every $1024$~rows, we keep the position of the first and of the last versioned row. With this information, it is possible to scan in tight loops between versioned records without performing any checks. 

Nevertheless, in Figure~\ref{figs:scan_over_versioned} we can see that this optimization can not defuse the problem entirely. With an increase is the number of versioned rows, we see a drastic increase in the runtime of the scan as well. Scanning a table that is completely versioned takes around $5$~times longer than scanning an unversioned table. This unversioned table essentially resembles the situation when scanning in a snapshot under heterogeneous processing. 

\subsection{Snapshotting Creation Cost}
\label{ssec:snapshotting_cost}

After inspecting the performance of transactions processing in form of latency in Section~\ref{ssec:latency}, throughput in Section~\ref{ssec:throughput}, and MVCC scan performance in Section~\ref{ssec:scan_performance}, let us inspect the cost of snapshot creation in AnKerDB. Due to our flexible system call~\smalltt{vm\_snapshot}, we are able to snapshot virtually at the granularity of individual columns.

\begin{figure}[!htb]
	\vspace{-0.2cm}
	\begin{center}
		\includegraphics[width=11cm, trim={0 0 0 0}, clip]{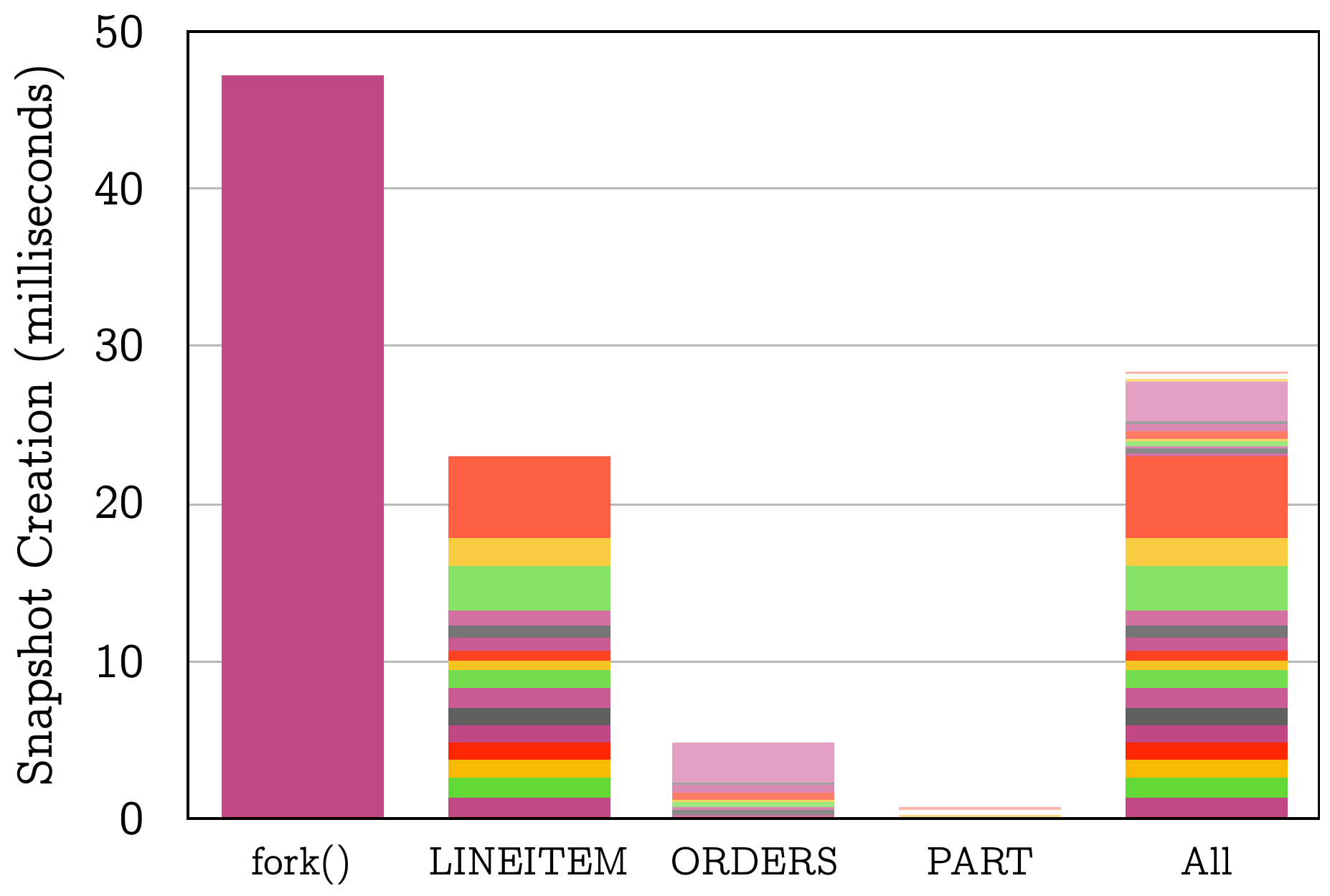}
	\end{center}
	\vspace{-0.5cm}
	\caption{\textbf{Snapshot creation cost} for the individual columns of \smalltt{LINEITEM}, \smalltt{ORDERS}, and \smalltt{PART} utilizing our system call~\smalltt{vm\_snapshot} in comparison with using \smalltt{fork}. }
	\label{figs:stacked_snapshot_cost}
\end{figure}

To demonstrate the benefit of this flexible approach, we present in Figure~\ref{figs:stacked_snapshot_cost} the cost of snapshotting the individual columns of the \smalltt{LINEITEM}, \smalltt{ORDERS}, and \smalltt{PART} table of the TPC-H benchmark inside of AnKerDB in form of stacked bars. Each layer in a bar resembles the cost of snapshotting a single column of the respective table. The bar~\textit{All} presents the cost of snapshotting all three tables. In comparison, we show the cost of forking the process in which AnKerDB is running using the system call~\smalltt{fork}. We make sure that when performing the \smalltt{fork}, the process is in the same state as when performing the snapshotting using \smalltt{vm\_snapshot}. At this point in time, the AnKerDB process has a size of~$5.2$GB in terms of virtual memory. 

As we can see in Figure~\ref{figs:stacked_snapshot_cost}, the cost of snapshotting individual columns of the TPC-H tables is negligibly cheap. Thus, if a transaction accesses only a portion of the attributes, the cost of preparing the snapshot stays as low as possible as well. Nevertheless, even when snapshotting all columns of all tables, our approach is considerably cheaper than using the \smalltt{fork}~system call. The problem of \smalltt{fork} is that the virtual memory of the entire process containing $5.2$GB of virtual memory is replicated. Besides the tables, which consume only around $1.5$GB of memory, this includes the used indexes, the version chains, the timestamp arrays, and various meta-data structures.

\subsection{Scaling}

Our system essentially implements parallelism on two layers: On the first layer, we parallelize OLTP and OLAP execution by maintaining the two processing components. On the second layer, we apply MVCC inside each component to ensure a high concurrency among transactions of a single type. In this regard, let us now investigate how well the design scales with the number of available threads. 

\begin{figure}[!htb]
	\vspace{-0.2cm}
	\begin{center}
		\includegraphics[width=11cm, trim={0 0 0 0}, clip]{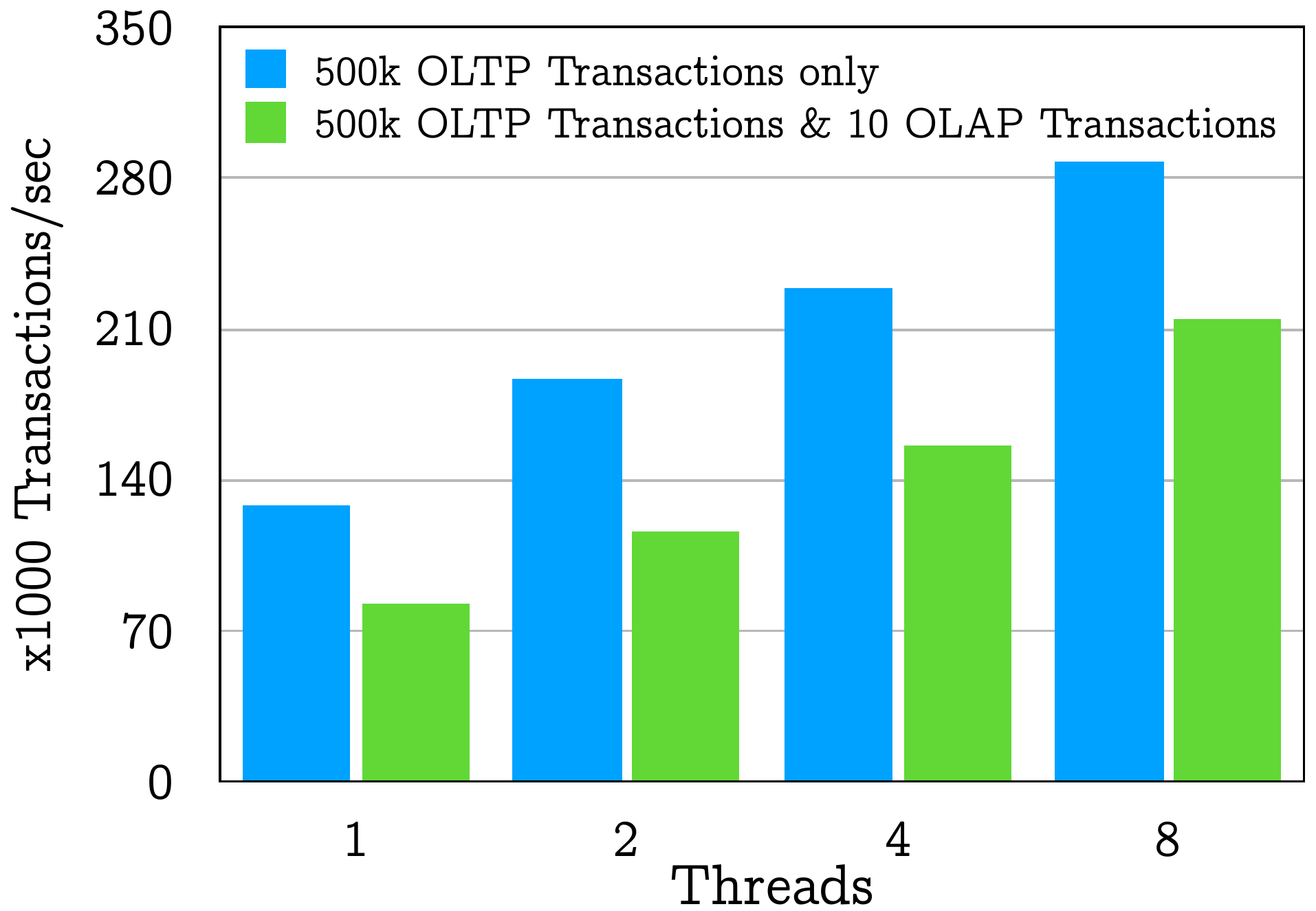}
	\end{center}
	\vspace{-0.5cm}
	\caption{\textbf{Scaling under heterogeneous processing (full serializability)}. We vary the number of threads from $1$ to $8$ and report the throughput under a pure OLTP workload as well as a mixed workload.}
	\label{figs:scaling}
\end{figure}

In Figure~\ref{figs:scaling}, we repeat the experiment measuring the throughput from Section~\ref{ssec:throughput} for heterogeneous processing and vary the number of available threads from $1$~thread to $8$~threads. As shown in Figure~\ref{figs:throughput}, we evaluate a pure OLTP workload consisting of $500{,}000$ transactions as well as a mixed workload that additionally runs $10$~OLAP transactions.

As we can see, the system scales sub-linear with the number of available threads. In comparison to single threaded execution, using $8$~threads results in a higher throughput of around $2.1$x for the OLTP workload and around $2.6$x for the mixed workload. The reason for this is that the commit phase validation, that is required for OLTP transactions to ensure full serializability, has to be partially sequential. For instance, a list of recently committed transactions, that must be mutex protected, is maintained to organize validation. Therefore, irrespective of our heterogeneous design, concurrent OLTP transaction processing under full serializability is limited by the validation phase.

\section{Conclusion}

In this work, we introduced AnKerDB, a transactional processing system implementing heterogeneous processing in combination with MVCC, which works hand in hand with a customized Linux kernel to enable snapshotting at a very high frequency. We have shown that a heterogeneous design powered by a lightweight snapshotting mechanism fits naturally to mixed OLTP/OLAP workloads and enhances the throughput of analytical transactions by factors $2$x to $4$x, as it enables very fast scans in tight loops. Besides, due to the flexibility of our custom system call~\smalltt{vm\_snapshot}, we are able to limit the snapshotting effort to those columns that are actually accessed by transactions, heavily reducing the snapshotting overhead in comparison to classical approaches.

{\small
\bibliographystyle{abbrv}
\bibliography{bibliography} 
}

\appendix

\section{VM\_SNAPSHOT Implementation Details}
\label{appendix:syscalls}

\noindent For the interested reader, the follow section provides a detailed description of the implementation details of \smalltt{vm\_snapshot}. 

\begin{minipage}{\linewidth}
\begin{small}
\begin{lstlisting}
void* vm_snapshot(void* dst_addr, 
                  void* src_addr, 
                  size_t length);
\end{lstlisting}
\label{listings:mapfileshared}
\end{small}
\end{minipage}

\begin{enumerate}

\item Check if the virtual memory area to snapshot in the range $[\smalltt{src\_addr},\smalltt{src\_addr}+\smalltt{length})$ is actually allocated. If no, the call fails with return value \smalltt{MAP\_FAILED} and sets \smalltt{errno} accordingly. 

\item Identify all VMAs that describe the virtual memory area $[\smalltt{src\_addr},\smalltt{src\_addr}+\smalltt{length})$. This might be one VMA or multiple ones. Let us call them in the following VMA$_0$ to VMA$_{n-1}$, if $n$ VMAs describe the area. 

\item It is possible that VMA$_0$ and VMA$_{n-1}$, the VMAs describing the borders of the virtual memory area, span larger than the area to replicate. This can be the case if virtual memory before \smalltt{src\_addr} or after $\smalltt{src\_addr}+ \smalltt{length}$ is currently allocated as well. In this case, we split VMA$_0$ and VMA$_{n-1}$ at \smalltt{src\_addr} respectively $\smalltt{src\_addr}+ \smalltt{length}$. If a split happens, we update VMA$_0$ and VMA$_{n-1}$ to the VMAs that now exactly match the borders of the region to replicate. 

\item If \smalltt{dst\_addr} is \smalltt{NULL}, reserve a new virtual memory area of size \smalltt{length} in the kernel, starting at address~\smalltt{dst\_addr}. If \smalltt{dst\_addr} is not \smalltt{NULL}, check whether 
$[\smalltt{dst\_addr},\smalltt{dst\_addr}+\smalltt{length})$ is already reserved and fail if not.

\item Iterate over VMA$_0$ to VMA$_{n-1}$. Let us refer to the current item as VMA$_i$. Further, let us define \smalltt{size}(VMA$_i$) as the size of the described virtual memory area and \smalltt{offset}(VMA$_i$) as the address of the described virtual memory area relative to \smalltt{src\_addr}. 
Now, we create an exact copy of VMA$_i$ and update the virtual memory area described by it to $[\smalltt{dst\_addr}$ + \smalltt{offset}(VMA$_i$), \smalltt{dst\_addr} + \smalltt{offset}(VMA$_i$) + \smalltt{size}(VMA$_i$)$)$.

\item Further, we check whether VMA$_i$ describes a shared or a private virtual memory area. If VMA$_i$ is shared, nothing more has to be done for this VMA. If VMA$_i$ is private, we additionally have to modify the page table, if there exist PTEs for the virtual memory area that VMA$_i$ is describing. In this case, we identify all $k$ PTEs, which relate to VMA$_i$, as PTE$_0$ to PTE$_{k-1}$.

\item Iterate over PTE$_0$ to PTE$_{k-1}$. Let us refer to the current item as PTE$_j$. If \smalltt{pageoffset}(PTE$_j$) returns the address of the mapped virtual page relative to \smalltt{src\_addr}, we create a copy of PTE$_j$ and update the start address of the mapped virtual page in the copy to \smalltt{dst\_addr} +  \smalltt{pageoffset}(PTE$_j$).
This step is necessary for private VMAs, as any write that is happening to the described virtual memory area results in a copy-on-write, that is handled with an anonymous physical page. As the information about the physical page is not present in the VMA but only in the corresponding PTE, we have to modify the page table in this case.

\end{enumerate}

\noindent After these steps, the virtual memory area $[\smalltt{dst\_addr},\smalltt{dst\_addr}+\smalltt{length})$ contains the snapshot and can be accessed.

\end{document}